\documentclass[aps,twocolumn,preprintnumbers,nofootinbib,superscriptaddress]{revtex4}
\usepackage{amsmath} \usepackage{graphicx} \usepackage{amsfonts}
\usepackage{array} \usepackage{amsthm} \usepackage{bm}
\usepackage{palatino} \usepackage{mathpazo} 
\usepackage{supertabular}

\usepackage[breaklinks]{hyperref}
\usepackage{color}

\usepackage{epstopdf}

\newcommand{\nn}{\nonumber}
\newcommand{\be}{\begin{equation}}
\newcommand{\ee}{\end{equation}}
\newcommand{\ba}{\begin{eqnarray}}
\newcommand{\ea}{\end{eqnarray}}
\newcommand{\bal}{\begin{align}}
\newcommand{\eal}{\end{align}}

\newcommand{\al}{\alpha}

\newcommand{\La}{\Lambda}
\newcommand{\bt}{\beta}

\newcommand{\ga}{\gamma}
\newcommand{\ro}{\rho}
\newcommand{\ep}{\epsilon}

\newcommand{\ta}{\theta}

\newcommand{\De}{\Delta}

\newcommand{\Om}{\Omega}

\newcommand{\de}{\delta}

\newcommand{\bw}{\begin{widetext}}
\newcommand{\ew}{\end{widetext}}

\begin{document}

\title{Cyclic and heteroclinic flows near general static spherically symmetric black holes}

\author{Ayyesha K. Ahmed}\email{ayyesha.kanwal@sns.nust.edu.pk}
\affiliation{Department of Mathematics, School of Natural
	Sciences (SNS), National University of Sciences and Technology
	(NUST), H-12, Islamabad, Pakistan}
\author{Mustapha Azreg-A\"{\i}nou}\email{azreg@baskent.edu.tr}
\affiliation{Engineering Faculty, Ba\c{s}kent University, Ba\u{g}l\i ca Campus, Ankara, Turkey}
\author{Mir Faizal}\email{mirfaizalmir@googlemail.com}
\affiliation{Department of Physics and Astronomy,
University of Lethbridge,  Alberta, T1K 3M4, Canada}
\affiliation{Department of Physics and Astronomy,  University of
Waterloo, Waterloo, Ontario, N2L 3G1, Canada}
\author{Mubasher Jamil}\email{mjamil@sns.nust.edu.pk}
\affiliation{Department of Mathematics, School of Natural
	Sciences (SNS), National University of Sciences and Technology
	(NUST), H-12, Islamabad, Pakistan}


\begin{abstract}
We investigate the Michel-type accretion onto a static spherically symmetric black hole. Using a Hamiltonian dynamical approach, we show that the standard method employed for tackling the accretion problem has masked some properties of the fluid flow. We determine new analytical solutions that are neither transonic nor supersonic as the fluid approaches the horizon(s); rather, they remain subsonic for all values of the radial coordinate. Moreover, the three velocity vanishes and the pressure diverges on the horizon(s), resulting in a flowout of the fluid under the effect of its own pressure. This is in favor of an earlier prediction that pressure-dominant regions form near the horizon. This result does not depend on the form of the metric and it applies to a neighborhood of any horizon where the time coordinate is timelike. For anti-de Sitter-like $\text{f}(R)$ black holes we discuss the stability of the critical flow and determine separatrix heteroclinic orbits. For de Sitter-like $\text{f}(R)$ black holes, we construct polytropic cyclic, non-homoclinic, physical flows connecting the two horizons. These flows become non-relativistic for Hamiltonian values higher than the critical value allowing for a good estimate of the proper period of the flow.
\end{abstract}

\maketitle



\section{Introduction}

General relativity is one of the most well tested theories in physics, however, there seem to be indications that
it might be modified at sufficiently large scales (as well as small scales). The most important indication of the modification of general relativity
comes from the
 observations made on the   Supernova type Ia (SN Ia) and Cosmic
Microwave Background (CMB) radiation  \cite{1,1d,d1}. These observations indicate that our universe is undergoing
 accelerated expansion. This could be explained by dark energy,
  and the vacuum energy in quantum field theories could have been used as a proposal for dark energy \cite{4,5}. However, the problem with this
  proposal is that the vacuum energy in quantum field theory is much more than the dark energy required to explain the present rate of expansion
  of the universe. There seem to be serious
  limitations on modifying  quantum field theories such that the vacuum energy is reduced to
  fit the amount of dark energy in the universe. In fact,  it has been argued
  that such modifications  will lead to   a violation
of the weak equivalence principle  \cite{6,7}.

The action for general relativity has also been  modified to explain the  accelerated expansion of the universe, and currently $\text{f}(R)$ gravity is one of the most well studied modifications of general relativity~\cite{1b,q2,q4,j1,j2,j3}. This is because the $\text{f}(R)$ gravity theories are known to produce an accelerated expansion of the universe \cite{1h,h1,j4}. Furthermore, if a cosmological constant exists, it will not have any measurable effect for most astrophysical phenomena \cite{4b,5b}.
However, the $\text{f}(R)$ gravity theories can have astrophysical consequences.
In fact, astrophysical consequences have also been used to constraint certain type of
$\text{f}(R)$ gravity models \cite{6b,7b}. So, it becomes both interesting and  important to study astrophysical phenomena using $\text{f}(R)$ gravity.
Several methods for the static spherically symmetric solutions in $\text{f}(R)$ gravity are studied in Refs \cite{FR1,FR3}. Regular black holes in $\text{f}(R)$ gravity are studied in Refs. \cite{FR2,FR4,FR6}. Myung discussed the stability of $\text{f}(R)$ black holes \cite{FR5}. Further, there are many applications of $\text{f}(R)$ gravity, e.g. gravity waves, brane models, effective equation approach, LHC test etc. \cite{L1,L2,L3}

An important astrophysical effect of black holes is that they tend to accrete matter, and such  accretion on a  black hole have been thoroughly studied~\cite{1u,W1,W2,2u}.
As the first studies of the accretion around a  black hole were done by Bondi in the Newtonian framework \cite{4u}, this effect is now known by the name of the Michel-type accretion. In his work, Bondi studied
the hydrodynamics of polytropic flow, and demonstrated that settling and transonic solutions exist for
  the gas accreting onto compact objects. The relativistic versions the Michel-type accretion
  have also been studied using  the steady state spherically symmetric flow of a test gas
around  a   black hole \cite{5u,6u}. It may be noted that the luminosity spectra and
the effect of an interstellar magnetic field in ionized gases \cite{7z},  the effect of radiative processes \cite{7z,9z,z9}, and the the effect of rotation \cite{8z} on accreting processes
 have also been studied. Recently, the Michel-type accretion of perfect fluids for a black hole in
 the presence of a cosmological constant has also been studied \cite{1t,t1,t}.  Jamil and collaborators studied the effects of phantom energy accretion onto static spherically symmetric black holes and the primordial black holes and found the masses of black holes to decrease and vanishing near the Big Rip \cite{j6,j7,j8,j9}.  The accretion on topologically charged black holes of the $\text{f}(R)$ theories and the
Einstein-Maxwell-Gauss-Bonnet black hole has also been investigated
by focusing on both inward and outward flows from the accretion disk
\cite{1j,2j}. Using the fact that data from the high-mass X-ray binary Cygnus X-1 has been used to constrain the values of the parameters for the $\text{f}(R)$ gravity theories \cite{t7}, in this paper,  we will rather analyze  some other aspects of  the Michel-type accretion for a black hole in a theory of $\text{f}(R)$ gravity.

The order of the paper is as follows. In Sec.~\ref{secGE} we discuss the general equations for spherical accretion including conservation laws for any static metric. We particularly show that the pressure of the perfect fluid for such spherically symmetric flows is, up to a sign, the Legendre transform of the energy density. This leads to a nice differential equation allowing the determination of the energy density, enthalpy, or pressure knowing one of the equations of state. In Sec.~\ref{secHS}, without restricting ourselves to a specific static black hole, we study the accretion phenomenon using the Hamiltonian dynamical system in the plane ($r,v$) where $r$ is the radial coordinate and $v$ is the three-dimensional speed of the fluid. We discuss sonic and non-sonic critical points for ordinary fluids as well as for non-ordinary matter. In Sec.~\ref{secFR} we write down the metric for static spherically symmetric black hole in a particular model of $\text{f}(R)$ gravity~\cite{1e} and discuss some of its properties. In Sec.~\ref{secITS} we study the isothermal fluid and various subcases. There we provide examples of new solutions among which critical flows and purely subsonic flows with vanishing speed and divergent pressure on the horizon as well as separatrix heteroclinic orbits by restricting the analysis to an $\text{f}(R)$ anti-de Sitter-like black hole. We also determine solutions that are purely supersonic and solution with transonic flows. We discuss the stability of some of these flows. In Sec.~\ref{secPTS} we apply the results of our Hamiltonian dynamical analysis to polytropic fluids. In Sec.~\ref{sec2m} we again consider the accretion of a polytropic fluid onto an $\text{f}(R)$ black hole solution where the function $\text{f}(R)$ is modeled by (a) Hu-Sawicki~\cite{m1} and (b) Starobinsky~\cite{m2} formulas. The last Section contains the conclusion and discussions of the above derivations.

Throughout the paper we have used the common relativistic notations. The chosen metric signature is $(-,+,+,+)$ and the geometric units  $G=c=1$.

\section{General equations for spherical accretion\label{secGE}}
In this section, in Sec.~\ref{secHS}, and in the first part of each of Sec.~\ref{secITS} and Sec.~\ref{secPTS} we consider any static spherically symmetric metric of the form
\begin{equation}\label{1}
ds^{2}= -fdt^2+\frac{dr^{2}}{f}+r^2(d\theta^2+\sin^2\theta d\phi^2),
\end{equation}
\emph{without specifying the form of the metric coefficient} $f$. Our results will apply to any black hole of that form and to any horizon in a neighborhood of which the time coordinate is timelike. In the second part of each of Sec.~\ref{secITS} and Sec.~\ref{secPTS} we consider some applications to an $\text{f}(R)$ anti-de Sitter-like, to Schwarzschild, and to an $\text{f}(R)$ de Sitter-like black holes.

In this section, we define the governing equations for spherical accretion. Here, we are considering the gas as a perfect fluid. We analyze the accretion rate and flow of a perfect fluid in $\text{f}(R)$ gravity. For this, we define the two basic laws of accretion i.e. particle conservation and energy conservation. We assume that the fluid is simple containing a single particle species; the fluid could be made of different particle species with low reactions rates or no reactions at all. Let $n$ be the baryon number density in the fluid rest frame and
\begin{equation}\label{velo}
u^{\mu}=dx^{\mu}/d\tau,
\end{equation}
be the intrinsic four velocity of the fluid where $\tau$ is the proper time. We define the particle flux or current density by $J^{\mu}=n u^{\mu}$. From the law of particle conservation, there will be no change in the number of particles i.e. neither particles are created nor destroyed. In other words, we say that for this system, the divergence of current density is conserved
\begin{equation}\label{6a}
\nabla_{\mu}J^{\mu}=\nabla_{\mu}(n u^{\mu})=0,
\end{equation}
where $\nabla_{\mu}$ is the covariant derivative. On the other hand, the stress-energy (SET) for a perfect fluid is given by
\begin{eqnarray}\label{7}
T^{\mu \nu}&=&(e+p)u^{\mu}u^{\nu}+p g^{\mu \nu},
\end{eqnarray}
where $e$ denotes the energy density and $p$ is the pressure. The Michel-type accretion is steady state and spherically symmetric \cite{1t,t1,t}, so all the physical quantities ($n,e,p,u^{\mu}$) and others that will be introduced later are functions of the radial coordinate $r$ only. Furthermore, we assume that the fluid is radially flowing in the equatorial plane $(\theta=\pi/2)$, therefore $u^{\theta}=0$ and $u^{\phi}=0$. For ease of notation we set $u^r=u$. Using the normalization condition $u^{\mu}u_{\mu}=-1$ and~\eqref{1}, we obtain,
\begin{equation}\label{9}
u_{t}= \pm\sqrt{f+u^2}.
\end{equation}

On the equatorial plane $(\theta=\pi/2)$, the continuity equation~(\ref{6a}) yields
\begin{eqnarray}\label{10}
\nabla_{\mu}(n u^{\mu})&=& \frac{1}{\sqrt{-g}}\partial_{\mu}(\sqrt{-g}n u^{\mu})\nonumber \\&=&
\frac{1}{r^{2}}\partial_{r}(r^{2}n u)=0.
\end{eqnarray}
or, upon integrating,
\begin{equation}\label{17}
r^{2}n u=C_{1},
\end{equation}
where $C_{1}$ is a constant of integration. This shows that, in a unit of proper time, the particle flux $\pi r^{2}n u$ through a sphere a radius $r$ remains constant for all $r$.

The thermodynamics of simple fluids is described by the two equations~\cite{Rezzolla}
\begin{equation}\label{t1}
dp=n(dh-Tds),\quad de=hdn+nTds,
\end{equation}
where $T$ is the temperature, $s$ is the specific entropy (entropy per particle), and
\begin{equation}\label{ent}
h=\frac{e+p}{n},
\end{equation}
is the specific enthalpy (enthalpy per particle)\footnote{If $m$ is the baryonic mass, then $\ro=mn$ is the mass density. Now, if $\mathfrak{h}=h/m$ and $\mathfrak{s}=s/m$ denote the enthalpy and entropy per unit mass, respectively, then $\ro\mathfrak{h}=nh$ and $\ro\mathfrak{s}=ns$. In terms of ($\mathfrak{h},\mathfrak{s},\ro$), Eqs.~\eqref{t1} and~\eqref{ent} take the forms $dp=n(d\mathfrak{h}-Td\mathfrak{s})$, $de=\mathfrak{h}d\ro+\ro Td\mathfrak{s}$, and $\mathfrak{h}=(e+p)/\ro$.}.

A theorem in relativistic hydrodynamics~\cite{Rezzolla,Gourgoulhon} states that the scalar $hu_{\mu}\xi^{\mu}$ is conserved along the trajectories of the fluid:
\begin{equation}\label{t3}
u^{\nu}\nabla_{\nu}(hu_{\mu}\xi^{\mu})=0,
\end{equation}
where $\xi^{\mu}$ is a Killing vector of spacetime generator of symmetry. In the special case we are considering in this work $\xi^{\mu}=(1,0,0,0)$ is timelike yielding
\begin{equation}\label{t4}
\partial_{r}(hu_{t})=0\quad\text{or}\quad h\sqrt{f+u^2}=C_2,
\end{equation}
where $C_2$ is a constant of integration. This equation can be derived directly upon evaluating
\begin{equation}\label{t5}
\nabla_{\mu}T^{\mu}{}_{t}=nu^{\mu}\nabla_{\mu}(hu_t)+\nabla_{t}(nh-e)=0,
\end{equation}
where we have used $T^{\mu}{}_{\nu}=nhu^{\mu}u_{\nu}+(nh-e)\de^{\mu}{}_{\nu}$. Since, the flow is stationary, any time derivative vanishes ($\nabla_{t}(nh-e)\equiv 0$), hence the result.

If the fluid had a uniform pressure, that is, if the fluid were not subject to acceleration, the specific enthalpy $h$ reduces to the particle mass $m$ and Eq.~\eqref{t3} reduces to $mu_{\mu}\xi^{\mu}=cst$ along the fluidlines. This is the well know energy conservation law which stems from the fact that the fluid flow is in this case geodesic. Now, if the pressure throughout the fluid is not uniform, acceleration develops through the fluid and the fluid flow becomes non-geodesic; the energy conservation equation $mu_{\mu}\xi^{\mu}=cst$, which is no longer valid, generalizes to its inertial equivalent~\cite{Rezzolla} $hu_{\mu}\xi^{\mu}=cst$ as expressed in Eqs.~\eqref{t3} and~\eqref{t4}.

It is well known that a perfect fluid~\eqref{7} is adiabatic; that is, the specific entropy is conserved along the evolution lines of the fluid ($u^{\mu}\nabla_{\mu}s=0$). This is easily established using the conservation of the SET, Eq.~\eqref{6a}, and the second equation in~\eqref{t1}. First, rewrite $T^{\mu\nu}$ as $nhu^{\mu}u^{\nu}+(nh-e)g^{\mu\nu}$, then project the conservation formula of the SET onto $u^{\mu}$
\begin{align}\label{t2}
u_{\nu}\nabla_{\mu}T^{\mu\nu}=&\ u_{\nu}\nabla_{\mu}[nhu^{\mu}u^{\nu}+(nh-e)g^{\mu\nu}]\nn\\
\quad =\ &u^{\mu}(h\nabla_{\mu}n-\nabla_{\mu}e)=-nTu^{\mu}\nabla_{\mu}s=0.
\end{align}
In the special case we are considering in this work where the fluid motion is radial, stationary (no dependence on time), and it conserves the spherical symmetry of the black hole, the latter equation reduces to $\partial_rs=0$ everywhere, that is, $s\equiv \text{const.}$. Thus, the motion of the fluid is isentropic and equations~\eqref{t1} reduce to
\begin{equation}\label{t1b}
dp=ndh,\quad de=hdn.
\end{equation}

Equations~\eqref{17} and~\eqref{t4} are the main equations that we will use to analyze the flow of a perfect fluid in the background of $\text{f}(R)$ black hole.

Another formula that will turn useful in the subsequent sections is the barotropic equation. Notice that the canonical form of the equation of state (EOS) of a simple fluid is $e=e(n,s)$~\cite{Gourgoulhon}. Since $s$ is constant, this reduces to the barotropic form
\begin{equation}\label{b1}
    e=F(n).
\end{equation}
From the second equation~\eqref{t1b} we have $h=de/dn$ yielding
\begin{equation}\label{b2}
    h=F'(n),
\end{equation}
where the prime denotes differentiation with respect to $n$. Now, the first equation~\eqref{t1b} yields $p'=nh'$ with $h=F'$ we obtain
\begin{equation}\label{b3}
p'=nF'',
\end{equation}
which we integrate by parts to derive
\begin{equation}\label{b4}
p=nF'-F.
\end{equation}
Here we identify, up to a sign, the Legendre transform of the energy density $F$. This conclusion is purely thermodynamic and it does not depend on the symmetric properties of the flow (presence of a timelike Killing vector and spherical symmetric flow); rather, it is valid for any isentropic flow ($s$ constant everywhere). The conclusion states that the pressure is the negative of the Legendre transform of the energy density and that an EOS of the form $p=G(n)$ is not independent of an EOS $e=F(n)$. The relationship between $F$ and $G$ can be derived upon integrating the first differential equation
\begin{equation}\label{b5}
nF'(n)-F(n)=G(n).
\end{equation}

In a locally inertial frame, the three-dimensional speed of sound $a$ is given by $a^2=(\partial p/\partial e)_s$~\cite{Weinberg}. Since the entropy $s$ is constant, this reduces to $a^2=dp/de$. Using~\eqref{t1b}, we derive a useful formula needed for the remaining sections
\begin{equation}\label{19}
a^2=\frac{dp}{de}=\frac{ndh}{hdn}\Rightarrow \frac{dh}{h}=a^{2}\frac{dn}{n}.
\end{equation}
Using~\eqref{b2}, this reduces to
\begin{equation}\label{19b}
a^2=\frac{ndh}{hdn}=\frac{n}{F'}F''=n(\ln F')'.
\end{equation}

Another useful formula is the three-velocity of a fluid element $v$ as measured by a locally static observer. Since the motion is radial in the plane $\ta=\pi/2$, we have $d\ta=d\phi=0$ and the metric~\eqref{1} implies the decomposition
\begin{equation*}
    ds^2=-(\sqrt{f}dt)^2+(dr/\sqrt{f})^2
\end{equation*}
in the standard special relativistic way~\cite{CT,Ellis} as seen by a locally static observer. The latter measures proper distances and proper times by $d\ell=dr/\sqrt{f}$ and $d\tau_0=\sqrt{f}dt$ corresponding to radial $dr$ and time $dt$ changes, respectively, and measures the three-velocity $v$ of the fluid element by
\begin{equation}\label{v0}
v\equiv \frac{d\ell}{d\tau_0}=\frac{dr/\sqrt{f}}{\sqrt{f}dt}.
\end{equation}
This yields
\begin{equation}\label{v}
v^2=\Big(\frac{u}{fu^t}\Big)^2=\frac{u^2}{u_t^2}=\frac{u^2}{f+u^2},
\end{equation}
where we have used $u^r=u=dr/d\tau$, $u^t=dt/d\tau$, $u_t=-fu^t$, and~\eqref{9}. This implies
\begin{equation}\label{v2}
u^2=\frac{fv^2}{1-v^2}\quad \text{ and }\quad u_t^2=\frac{f}{1-v^2},
\end{equation}
and~\eqref{17} becomes
\begin{equation}\label{v3}
\frac{r^4n^2fv^2}{1-v^2}=C_1^2.
\end{equation}

In relativistic hydrodynamics one usually derives the above formulas on considering the woldlines of a fluid element and that of a locally static observer. If $\mathbf{u}$ and $\mathbf{u_0}$ are the respective four-velocities, we have~\cite{Gourgoulhon,faces}
\begin{equation}\label{v4}
\mathbf{u}=\Gamma (\mathbf{u_0}+\mathbf{U})\qquad (\text{with }\ \mathbf{u_0}\cdot \mathbf{U}=0),
\end{equation}
where $\mathbf{U}$ is the relative four-velocity, that is, the velocity of the observer attached to the fluid element relative to the locally static observer with the property $\mathbf{u_0}\cdot \mathbf{U}=0$, where the dot represents the scalar product with respect to the metric~\eqref{1}. $\Gamma$ is the Lorentz factor $\Gamma\equiv -\mathbf{u_0}\cdot \mathbf{u}=d\tau_0/d\tau$~\cite{Gourgoulhon,faces}. In the case of radial motion in the $\ta =\pi/2$ plane, we have
\begin{align}\label{v5}
&\mathbf{u}=(u^t,u,0,0)=u^t\partial_t+u\partial_r,\nn\\
&\mathbf{u_0}=(1/\sqrt{f},0,0,0)=\partial_t/\sqrt{f},\\
&\mathbf{U}=(0,V^r,0,0)=V^r\partial_r.\nn
\end{align}
Here $u^t$ and $u=u^r$ are as defined in~\eqref{velo} and $V^r=dr/d\tau_0=\sqrt{f}v$. Since $\partial_r$ is not a unit four vector, rather it is $v$, and not $V^r$, the three velocity that the locally static physical observer, who uses the orthonormal basis ($\partial_t/\sqrt{f},\sqrt{f}\partial_r,\partial_{\theta}/r,\partial_{\phi}/r$), measures. Squaring~\eqref{v4} we obtain
\begin{equation}\label{v6}
    \Gamma =\frac{1}{\sqrt{1-\mathbf{U}\cdot \mathbf{U}}}=\frac{1}{\sqrt{1-v^2}},
\end{equation}
where we have used $\mathbf{U}\cdot \mathbf{U}=g_{rr}V^rV^r=v^2$ in the last expression. The expressions~\eqref{v2} are rederived from~\eqref{v4}, \eqref{v5}, and~\eqref{v6}.

All the above expressions remain valid for an observer outside the horizon, more precisely, for an observer where the time coordinate is timelike. We define the value $v_h$ of $v$ on the horizon(s) $r_h$ as the limit of the continuous three velocity field $v(r)$ as $r$ approaches $r_h$ from within the region where the time coordinate is timelike ($f>0$):
\begin{equation}\label{v7}
v_h=\lim _{r\underset{(f>0)}{\to}r_h}v(r).
\end{equation}

\section{Hamiltonian systems\label{secHS}}

We have derived two integrals of motion ($C_1,C_2$) given in~\eqref{17} and~\eqref{t4}. Either of these integrals, or any combination of them, can be used as a Hamiltonian for the fluid flow. The simplest Hamiltonian system has one degree of freedom, in which case the Hamiltonian $\mathcal{H}$ is a two-variable function ($x,y$). Let $\mathcal{H}$ be the square of the lhs of~\eqref{t4}:
\begin{equation}\label{h1}
\mathcal{H}=h^2(f+u^2).
\end{equation}
Now, we need to fix the two dynamical variables ($x,y$) on which $\mathcal{H}$ depends and the time variable $\bar{t}$ of the Hamiltonian dynamical system. There are different ways to fix the dynamical variables; one may choose ($x,y$) to be ($r,u$)~\cite{t}, ($r,v^2$)~\cite{t}, ($r,n$)~\cite{CS}, ($r,h$), or even ($r,p$). The time variable $\bar{t}$ for the dynamical system is any variable on which $\mathcal{H}$~\eqref{h1} does not depend explicitly so that the dynamical system is autonomous.

In Sec.~\ref{secGE} we have seen that, under the symmetry requirements of the problem, $h$ is an explicit function of the baryon number density $n$ only; this applies to the pressure $p$ too. So, if ($x,y$) are chosen to be ($r,h$) (resp. ($r,p$)), the Hamiltonian~\eqref{h1} takes the form
\begin{equation}\label{h2}
\mathcal{H}=h(n)^2\Big[f(r)+\frac{C_1^2}{r^4n^2}\Big]\qquad (C_1^2>0),
\end{equation}
where we have used~\eqref{17} (resp. $\mathcal{H}=h(p)^2\big[f(r)+\tfrac{C_1^2}{r^4n(p)^2}\big]$).

This conclusion does not extend to other dynamical variables, that is, if one chooses ($x,y$) to be, say, ($r,v$), it is not true to assume $h=h(r)$ or $h=h(v)$, for, by~\eqref{17} and~\eqref{v2}, $n$ is a function of ($r,v$) and so is $h$. With $h=h(r,v)$, the Hamiltonian~\eqref{h1} of the dynamical system reads
\begin{equation}\label{h3}
\mathcal{H}(r,v)=\frac{h(r,v)^2f(r)}{1-v^2},
\end{equation}
where we have used~\eqref{v2} to eliminate $u^2$ from~\eqref{h1}. We have thus fixed the dynamical variable to be $(r,v)$. No use has been made of~\eqref{17} to derive~\eqref{h3}; use of it will be made in the derivation of the critical points (CPs), particularly, of the sonic points.

From now on, partial derivatives will be denoted as $\partial f/\partial x=f_{,x}$.

\subsection{Sonic points}

In the remaining part of this section, we assume that the parametric Hamiltonian of the dynamical system is given by~\eqref{h3}. In this section we use~\eqref{h3} to derive the CPs of the dynamical system and derive them in the Appendix B~\ref{secab} using~\eqref{h2}.

With $\mathcal{H}$ given by~\eqref{h3}, the dynamical system reads
\begin{equation}\label{h4}
\dot{r}=\mathcal{H}_{,v}\,,  \quad\quad \dot{v}=-\mathcal{H}_{,r}.
\end{equation}
(here the dot denotes the $\bar{t}$ derivative). In~\eqref{h4} it is understood that $r$ is kept constant when performing the partial differentiation with respect to $v$ in $\mathcal{H}_{,v}$ and that $v$ is kept constant when performing the partial differentiation with respect to $r$ in $\mathcal{H}_{,r}$. We will keep using this simple notation in the subsequent steps of this section. The CPs of the dynamical system are the points ($r_c,v_c$) where the rhs's in~\eqref{h4} are zero. Evaluating the rhs's we find
\begin{align}
\label{h5a}&\mathcal{H}_{,v}=\frac{2fh^2v}{(1-v^2)^2}\Big[1+\frac{1-v^2}{v}~(\ln h)_{,v}\Big],\\
\label{h5b}&\mathcal{H}_{,r}=\frac{h^2}{1-v^2}\big[f_{,r}+2f~(\ln h)_{,r}\big].
\end{align}

The rightmost formula in~\eqref{19} yields
\begin{equation}\label{h6}
(\ln h)_{,v}=a^2(\ln n)_{,v}\quad\text{ and }\quad (\ln h)_{,r}=a^2(\ln n)_{,r}.
\end{equation}
Now, using~\eqref{v3} we see that if $r$ is kept constant we have the equation $nv/\sqrt{1-v^2}=\text{const.}$ which upon differentiating with respect to $v$ we obtain
\begin{equation}\label{h7}
(\ln n)_{,v}=-\frac{1}{v(1-v^2)}\Rightarrow (\ln h)_{,v}=-\frac{a^2}{v(1-v^2)};
\end{equation}
and if $v$ is kept constant we have the equation $r^2n\sqrt{f}=\text{const.}$ which upon differentiating with respect to $r$ we obtain
\begin{equation}\label{h8}
(\ln n)_{,r}=-\frac{4+r(\ln f)_{,r}}{2r}\Rightarrow (\ln h)_{,r}=-\frac{a^2[4+r(\ln f)_{,r}]}{2r}.
\end{equation}
Finally, the system~\eqref{h4} reads
\begin{align}
\label{h9a}&\dot{r}=\frac{2fh^2}{v(1-v^2)^2}~(v^2-a^2),\\
\label{h9b}&\dot{v}=-\frac{h^2}{r(1-v^2)}[rf_{,r}(1-a^2)-4fa^2].
\end{align}
Let us assume that $h$ is never zero and finite (the same applies to $n$). The rhs's vanish if
\begin{equation}\label{cp1}
v_c^2=a_c^2\quad\text{ and }\quad r_c(1-a_c^2)f_{c,r_c}=4f_ca_c^2,
\end{equation}
where $f_c=f(r)|_{r=c}$ and $f_{c,r_c}=f_{,r}|_{r=c}$. The second equation expresses the speed of sound at the CP, $a_c^2$, in terms of $r_c$
\begin{equation}\label{cp2}
    a_c^2=\frac{r_cf_{c,r_c}}{r_cf_{c,r_c}+4f_c},
\end{equation}
which will allow to determine $r_c$ once the EOS $a^2=dp/de$ [or $e=F(n)$] is known. The remaining needed ingredient is a simplified expression for $n/n_c$. If we write the constant $C_1^2$ in~\eqref{v3} as
\begin{equation}\label{b9}
    C_1^2=r_c^4n_c^2v_c^2~\frac{f_c}{1-v_c^2}=r_c^4n_c^2v_c^2~\frac{r_cf_{c,r_c}}{4v_c^2}=\frac{r_c^5n_c^2f_{c,r_c}}{4},
\end{equation}
where we have used~\eqref{cp1}. Using this in~\eqref{v3} we obtain
\begin{equation}\label{b10}
\Big(\frac{n}{n_c}\Big)^{2}=\frac{r_c^5f_{c,r_c}}{4}~\frac{1-v^2}{r^4fv^2}.
\end{equation}
As we shall see in the subsequent sections, there will be two types of fluid flow approaching the horizon, in the one type the speed $v$ vanishes and in the other one the speed approaches that of light in such a way that the ratio $(1-v^2)/f$ may remain finite. In the former type of motion, the number density $n$ diverges on the horizon independently of the expression of $f$.

An expression for $u_c^2$ is derived upon substituting~\eqref{cp1} into~\eqref{v2}, then making use of~\eqref{cp2}
\begin{equation}\label{cp4}
    u_c^2=\frac{fa_c^2}{1-a_c^2}=\frac{r_cf_{c,r_c}}{4}.
\end{equation}

Another sonic CP is the point corresponding to $f_c=0$ and $a_c^2=1$. But the roots of $f_c=0$ may coincide with the horizons $r_h$ of the black hole. This implies that the fluid becomes ultra-stiff as it approaches the horizon where $r_c=r_h$ (the fluid is not necessarily ultra-stiff for all $r$). This conclusion does not apply to $\text{f}(R)$ gravity only; rather, to any static spherically symmetric metric of the form~\eqref{1}. To the best of our knowledge, this result has not been announced elsewhere. Now, by~\eqref{v3}, since $f_c=0$ we must necessarily have $v_c^2=1$.  This point, however, may fail to behave as a CP in the mathematical sense, for the rhs's of~\eqref{h9a} and~\eqref{h9b} may become undetermined or may have nonzero values there. This point, ($r=r_h,v=1$), may behave as a focus point as we shall see in the next section.

\subsection{Non-sonic critical points}
From~\eqref{h9a}, we see that $f_c=0$ and $f_{c,r_c}=0$ may lead to a non-sonic CP. However, this CP would be a double root of $f=0$, which is out of the scope of this paper where we only consider non-extremal black holes.

Another obvious CP, which lies within the scope of $\text{f}(R)$ gravity, corresponds to $h(r_c)=0$~\eqref{h9a} and~\eqref{h9b}. This is not possible for ordinary matter but is the case for non-ordinary matter with negative pressure. When this is the case, $h$ may vanish at some point with no special constraint on $v^2$ and $a^2$. This means that for non-ordinary fluids, the flow may not become transonic at all.  We will not pursue this discussion here, for it is out of the scope of this work. In the next section, however, we will pursue this discussion for ordinary matter where it is generally admitted that ``\texttt{the flow must be supersonic at the horizon, though it is necessarily subsonic at a large distance}"~\cite{Sandip}. We will explicitly show, through physical solutions, the existence of subsonic flow for all values of the radial coordinate. Moreover, the speed of the flow vanishes as the fluid approaches the horizon, so the flow does not necessary become supersonic nor transonic near the horizon~\cite{NT1,NT2}. Our conclusion remains true even for the Schwarzschild black hole. We believe that the use of standard methods for tackling the accretion problems has masked many features of them.

The conclusions made in this section, concerning the sonic CP [from~\eqref{h9a} to~\eqref{cp4}], do not apply to $\text{f}(R)$ gravity only, for we have not fixed the form of the metric coefficient $f$ yet; they apply to any static metric with $g_{tt}=-1/g_{rr}$ and $g_{\ta\ta}=r^2$.

Applications are given in the following sections where we consider three models of $\text{f}(R)$ gravity.

\section{Black hole in $\text{f}(R)$ gravity\label{secFR}}
Recently, an interesting model of $\text{f}(R)$ gravity has been proposed \cite{1e}, and the
  motion of test particles around a  black hole   in this theory   has been investigated.
The Lagrangian for this model of $\text{f}(R)$ theory is given by \cite{1e},
\begin{eqnarray}
 \text{$\text{f}(R)$} = R + \Lambda + \frac{R + \Lambda}{ d^2(6 \alpha^2)^{-1}R + 2 \alpha ^{-1}}\ln \frac{R + \Lambda}{ R_c},
\end{eqnarray}
where $\Lambda$ is the cosmological constant, $R_c$ is a constant of integration\footnote{$R_c$ is merely a constant of integration which is used to balance the dimensions of $R$. Its value, which ``\texttt{is not sensitive to the SNIa data"}~\cite{2e}, is not known by any physical theory and can only be determined using astronomical constraints as suggested by Safari and Rahvar~\cite{2e}.}, and
$\alpha, d$ are free parameters of this theory. The limit that is relevant for
astrophysical scale  corresponds to   $R\gg \Lambda$ and $d^2(6 \alpha^2)^{-1}R \gg 2\alpha$. In this limit, we obtain $\text{$\text{f}(R)$} = R + \Lambda +  d^2(6 \alpha^2)^{-1}R \ln \frac{R}{ R_c}$. The limit that is relevant to the cosmological scale is $R \sim Rd^2(6 \alpha^2)^{-1} \sim \Lambda $ yielding $\text{$\text{f}(R)$} = R + \Lambda$. This limit constrains the accelerating expansion \cite{2e}. It is useful to introduce a parameter $\beta = \alpha/d$ in terms of which both limits of the theory can be studied \cite{1e}. In this theory, the  metric for a spherically symmetric black hole with mass $M$ takes the form,
\begin{multline}\label{1met}
ds^{2}= -fdt^2+\frac{dr^{2}}{f}+r^2(d\theta^2+\sin^2\theta d\phi^2)\\
\text{with }\quad f\equiv 1-\frac{2M}{r}+\beta r-\frac{\Lambda r^{2}}{3}.\quad\qquad\qquad
\end{multline}
If $\Lambda=0$, (\ref{1met}) reduces to a special case of Kiselev black hole \cite{Kis,Azka} and if $\beta=0$, (\ref{1met}) reduces to Schwarzschild--de-Sitter or Schwarzschild--anti-de-Sitter black hole.

The present model of $\text{f}(R)$ can explain the flat rotation curve of
galaxies, consistent with solar system tests and also explains the
pioneer anomaly/acceleration. For details concerning the motivation
for this particular model of $\text{f}(R)$ theory, we refer the reader to the
original work by Saffari and Rahvar~\cite{2e}. Of course the
present analysis can also be done for other $\text{f}(R)$ black holes such as
Eq. 32 of Ref~\cite{CDM} and will be reported elsewhere. However
due to the generality of our work, further analysis will be trivial as was the case with $\text{f}(T)$ gravity black holes~\cite{FTG}.

It is well-known that $\text{f}(R)$ theory has a representation equivalent to
a particular class of scalar-tensor (ST) theories namely, the
Brans-Dicke (BD) theory i.e. a scalar field being non-minimally coupled
to gravity or curvature with vanishing kinetic term of the scalar
field. This description holds for both metric and palatini $\text{f}(R)$
theories~\cite{hair1,hair2}. Furthermore, the no-hair theorem for black holes in a general ST theory suggests that
the Schwarzschild solution is the only asymptotically flat, exterior,
vacuum, static and spherically symmetric solution to ST theory~\cite{hair3}. However, it does not rule out the existence of non-asymptotically flat ST black holes without hair. For instance, the Reissner-Nordstr\"om Anti-de Sitter
kind of topological black holes are derived in BD-Maxwell ST theory~\cite{hair4}. In the same context, we study a non-asymptotically flat $\text{f}(R)$ black hole.

The roots of $f=0$, or equivalently, the roots of $P=0$, where $P\equiv 3rf=-\Lambda r^{3}+3\bt r^2+3r-6M$ is a polynomial of degree 3, determine all possible horizons of~\eqref{1met}. If $\La>0$, the equation $P=0$ has always some negative root, which we ignore because of the physical singularity at $r=0$, and it may have two positive roots or a double positive root depending on the values of its coefficients. These two positive roots, if any, determine the event and cosmological horizons. In this case, the fluid flow would be confined in the space region enclosed by the two horizons. If there are no positive roots, the metric coefficient $g_{tt}$ is positive for all $r>0$; this case is not interesting.

We will be interested in the cases where the positive roots of $P=0$ are single. Assuming $\La <0$ (anti-de Sitter-like black hole) and $\bt\geq0$, then if $\bt^2>-\La$, $P=0$ has either two negative roots and one positive root or one double negative root and one positive root; if $0\leq \bt^2\leq -\La$, $P=0$ has one single positive root. On converting the polynomial $P(r)$ into the Weierstrass polynomial $w(z)\equiv 4z^3-g_2z-g_3$ by the transformation $r=z+\bt/\La$, we can parameterize the roots of $P=0$ based on the parametrization of the roots of $w(z)$ as given in the Appendix A~\ref{secaa}~\cite{LP}. The horizon is given by
\begin{equation}\label{3a}
r_h=\frac{\bt}{\La}+\sqrt{\frac{g_2}{3}}\cos\Big(\frac{\eta}{3}\Big),
\end{equation}
if $P=0$ has at least two real roots;
\begin{multline}\label{3b}
r_h=\frac{\bt}{\La}+\frac{1}{2\cdot 9^{1/3}}[(9 g_3+ \sqrt{3} \sqrt{-\Delta })^{1/3}\\+(9 g_3-\sqrt{3} \sqrt{-\Delta })^{1/3}],
\end{multline}
if $P=0$ has only one real root. Here $g_2$ and $g_3$ are defined by
\begin{equation}\label{3c}
g_2=\frac{12 (\beta ^2+\Lambda )}{\Lambda ^2},\quad g_3=\frac{4 (2 \beta ^3+3 \beta  \Lambda -6 M \Lambda ^2)}{\Lambda^3},
\end{equation}
and $\De$ and the angle $0\leq \eta\leq \pi$ are defined as in Eqs.~\eqref{df2} and~\eqref{3.4}, respectively.

Now, assuming $\La >0$ (de Sitter-like black hole) and $\bt\geq0$, $P=0$ has always one negative root and will have two positive roots, corresponding to an event horizon $r_{eh}$ and a cosmological horizon $r_{ch}>r_{eh}$ if $2(\bt^2+\La)r_+>6M\La-\bt$ where $r_+$ is the positive root of $P'(r)=0$. When this the case, the roots are given
\begin{multline}\label{3d}
r_{ch}=\frac{\bt}{\La}+\sqrt{\frac{g_2}{3}}\cos\Big(\frac{\eta}{3}\Big),\\ r_{eh}=\frac{\bt}{\La}-\sqrt{\frac{g_2}{3}}\cos\Big(\frac{\pi+\eta}{3}\Big),
\end{multline}
where $g_2$ and $g_3$ are defined by~\eqref{3c}. $\De$ and the angle $0\leq \eta\leq \pi$ are defined as in Eqs.~\eqref{df2} and~\eqref{3.4}, respectively. To have a common notation with the case $\La<0$, we will for short denote $r_{eh}$ and $r_{ch}$ by $r_{h}$.

The scalar invariants $R$, $R^{\mu \nu}R_{\mu \nu}$, and $R^{\mu \nu \sigma \rho}R_{\mu \nu \sigma \rho}$ are given by
\begin{eqnarray}\label{6}
I_{1}&=&R=\frac{6\beta}{r}-4\Lambda,\\
I_{2}&=&R^{\mu \nu}R_{\mu \nu}=\frac{2(5\beta^{2}-6r\beta \Lambda+2r^{2}\Lambda^{2})}{r^{2}},\\
I_{3}&=&R^{\mu \nu \sigma \rho}R_{\mu \nu \sigma \rho}=\frac{48M^{2}}{r^{6}}+\frac{8\beta^{2}}{r^{2}}-\frac{8\beta\Lambda}{r}+\frac{8\Lambda^{2}}{3},
\end{eqnarray}
which reduce to the Schwarzschild values $I_{1}=I_{2}=0$ and $I_{3}=48M^{2}/r^{6}$ if $\beta=\Lambda=0$. Clearly $r=0$ is the curvature singularity, which is not removable.

\section{Isothermal test fluids\label{secITS}}

Isothermal flow is often referred to the fluid flowing at a constant temperature. In other words, we can say that the sound speed of the accretion flow remains constant throughout the accretion process. This ensures that the sound speed of accretion flow at any radii is always equivalent to the sound speed at sonic point \cite{Review}. Here our system is adiabatic, so it is more likely that the flow of our fluid is isothermal in nature. Therefore, in this section we find the general solution to the isothermal equation of state of the form $p=ke$, that is of the form $p=kF(n)$~\eqref{b1} with $G(n)=kF(n)$~\eqref{b5}. Here $k$ is the state parameter constrained by $(0<k\leq1)$~\cite{t1}.
Generally, the adiabatic sound speed is defined as $a^{2}= {dp}/{de}$. So by  comparing the adiabatic sound speed to the equation of state, we find $a^{2}=k$.

The differential equation~\eqref{b5} reads
\begin{equation}\label{b6}
nF'(n)-F(n)=kF(n),
\end{equation}
yielding
\begin{equation}\label{b7}
    e=F=\frac{e_c}{n_c^{k+1}}\,n^{k+1},
\end{equation}
where we have chosen the constant of integration\footnote{This constant, $e_c/n_c^{k+1}$, in~\eqref{b7} could have been chosen $e_{\infty}/n_{\infty}^{k+1}$ or $e_0/n_0^{k+1}$ where ($e_0,n_0$) are any reference (energy density, number density).} so that~\eqref{ent} and~\eqref{b2} lead to the same expression for $h$
\begin{equation}\label{b8}
    h=\frac{(k+1)e_c}{n_c^{k+1}}\,n^{k}=\frac{(k+1)e_c}{n_c}\Big(\frac{n}{n_c}\Big)^{k}.
\end{equation}
Now, setting
\begin{equation*}
    K=\Big(\frac{r_c^5f_{c,r_c}}{4}\Big)^{k}\Big(\frac{(k+1)e_c}{n_c}\Big)^{2}=\text{const.},
\end{equation*}
and using~\eqref{b10} we simplify $h(r,v)^2$ by
\begin{equation}\label{b11}
    h^2=K\Big(\frac{1-v^2}{v^2r^4f}\Big)^{k}.
\end{equation}
Upon performing the transformation $\bar{t}\to K\bar{t}$ and $\mathcal{H}\to \mathcal{H}/K$, the constant $K$ gets absorbed in a redefinition of the time $\bar{t}$. Using~\eqref{b11}, the new Hamiltonian $\mathcal{H}$ and the dynamical system~\eqref{h9a}, \eqref{h9b} read
\begin{align}\label{nds}
&\mathcal{H}(r,v)=\frac{f}{1-v^2}\Big(\frac{1-v^2}{v^2r^4f}\Big)^{k}=\frac{f^{1-k}}{(1-v^2)^{1-k} v^{2 k} r^{4 k}},\nn\\
&\dot{r}=\frac{2(v^2-a^2)f}{v(1-v^2)^2}\Big(\frac{1-v^2}{v^2r^4f}\Big)^{k},\\
&\dot{v}=-\frac{1}{r(1-v^2)}\Big(\frac{1-v^2}{v^2r^4f}\Big)^{k}[rf_{,r}(1-a^2)-4fa^2],\nn
\end{align}
where the dot denotes differentiation with respect to the new time $\bar{t}$.

For a subsequent physical discussion we need an expression for the pressure. With $p=ke$, we obtain upon substituting~\eqref{b10} into~\eqref{b7}
\begin{equation}\label{pr1}
p\propto \Big(\frac{1-v^2}{v^2r^4f}\Big)^{\frac{k+1}{2}}.
\end{equation}
Since the Hamiltonian~\eqref{nds} remains constant on a solution curve, if the latter approaches the horizon (any horizon) from within the region where $t$ is timelike, $f$ approaches 0, and so the speed $v$ must either approach 1 or 0 so that the Hamiltonian retains the same constant value (otherwise, the Hamiltonian would always assume a 0 value on the horizon regardless its constant value elsewhere). In former case ($v\to 1$), the pressure~\eqref{pr1} may remain finite in a neighborhood of the horizon. In the latter case ($v\to 0$), the pressure diverges as the solution curve approaches the horizon. This is a very general conclusion which holds for any metric coefficient $f$ and any horizon of the black hole. If the latter is of de Sitter type ($\La >0$), a pressure-dominant region may form near both the event and cosmological horizons. This is in favor of a proposal that a pressure-dominant region would form near the horizon~\cite{Shapiro}.

If $f(r)=0$ has a single root as $r$ approaches $r_h$ (corresponding to an event, a cosmological, or any horizon in a neighborhood of which $t$ is timelike), which is our case, then, in the latter case ($v\to 0$), as the curve approaches the horizon $f\sim (r-r_h)$ and $v^{2k}\sim f^{1-k}$, thus $v^2\sim  (r-r_h)^{(1-k)/k}$. Using this in~\eqref{pr1} we see that the pressure diverges, as the curve approaches the horizon, as
\begin{equation}\label{pr2}
    p\sim (r-r_h)^{-\frac{k+1}{2k}}.
\end{equation}
If $r_h$ is a double root of $f=0$, we obtain
\begin{equation*}\label{pr2b}
    p\sim (r-r_h)^{-\frac{k+1}{k}}.
\end{equation*}

Before we proceed further let us see what the constraints on $k$ to have a physical flow are. Along a solution curve, the Hamiltonian of the dynamical system~\eqref{nds} is constant [where the constant is proportional to $C_2$~\eqref{t4}]. A global flow solution that extends to spatial infinity corresponds to
\begin{equation}\label{gf1}
    v\simeq v_1r^{-\al}+v_{\infty}\quad\text{ as }\quad r\to\infty,
\end{equation}
where ($\al > 0,v_1,|v_{\infty}|\leq 1$) are constants. Inserting this in the Hamiltonian~\eqref{nds} reduces to

\begin{equation}\label{gf2}
\mathcal{H}\simeq\left\{
  \begin{array}{ll}
    \text{(a):}\quad\frac{f^{1-k}}{r^{4k}}, & \hbox{(if $0<|v_{\infty}|< 1$);} \\
    \text{(b):}\quad\frac{f^{1-k}}{r^{(4-2\al)k}}, & \hbox{(if $v_{\infty}= 0$);}\\
    \text{(c):}\quad\frac{f^{1-k}}{r^{(4+\al)k-\al}}, & \hbox{(if $|v_{\infty}|= 1$);}
  \end{array}
\right.
\end{equation}
Using the metric~\eqref{1met}, each case splits into two subcases as follows.
\begin{equation}\label{gf3}
\text{(a)}\Rightarrow\left\{
  \begin{array}{ll}
    \text{(a1):}\quad\mathcal{H}\simeq r^{2-6k}, & \hbox{(if $\La\neq 0$);} \\
    \text{(a2):}\quad\mathcal{H}\simeq r^{1-5k}, & \hbox{(if $\La = 0,\bt\neq 0$).}
  \end{array}
\right.
\end{equation}
Since $\mathcal{H}$ is constant along a solution curve we must have $k=1/3$ ($\La\neq 0$) and $k=1/5$ ($\La = 0,\bt\neq 0$), respectively. These are the only possibilities allowing for a fluid flow with a nonvanishing, nonrelativistic three-dimensional speed.
\begin{equation}\label{gf4}
\text{(b)}\Rightarrow\left\{
  \begin{array}{ll}
    \text{(b1):}\quad\mathcal{H}\simeq r^{2-6k+2\al k}, & \hbox{(if $\La\neq 0$);} \\
    \text{(b2):}\quad\mathcal{H}\simeq r^{1-5k+2\al k}, & \hbox{(if $\La = 0,\bt\neq 0$).}
  \end{array}
\right.
\end{equation}
Thus, for ordinary fluids we deduce
\begin{align}
\label{gf5a}&\text{(b1):}\quad\tfrac{1}{3}<k<1\quad\text{ and }\quad 0<\al\leq 2,\\
\label{gf5b}&\text{(b2):}\quad\tfrac{1}{5}<k<1\quad\text{ and }\quad 0<\al\leq 2,
\end{align}
and for non-ordinary fluids ($-1\leq k<0$) we deduce
\begin{align}
\label{gf5c}&\text{(b1):}\quad -1\leq k<0\quad\text{ and }\quad \al\geq 4,\\
\label{gf5d}&\text{(b2):}\quad -1\leq k<0\quad\text{ and }\quad \al\geq 3.
\end{align}
On comparing the leading terms in the expansion~\eqref{gf1}, we see that the fluid flow for ordinary matter is faster at spatial infinity than it is for non-ordinary matter.
\begin{equation}\label{gf6}
\hspace{-2mm}\text{(c)}\Rightarrow\left\{
  \begin{array}{ll}
    \text{(c1):}\quad\mathcal{H}\simeq r^{2-6k+\al-\al k}, & \hbox{(if $\La\neq 0$);} \\
    \text{(c2):}\quad\mathcal{H}\simeq r^{1-5k+\al-\al k}, & \hbox{(if $\La = 0,\bt\neq 0$).}
  \end{array}
\right.
\end{equation}
Thus, for ordinary fluids we deduce
\begin{align}
\label{gf7a}&\text{(c1):}\quad\tfrac{1}{3}< k<1\quad\text{ and }\quad \al =\tfrac{2(3k-1)}{1-k} >0,\\
\label{gf7b}&\text{(c2):}\quad\tfrac{1}{5}< k<1\quad\text{ and }\quad \al =\tfrac{5k-1}{1-k} > 0,
\end{align}
while for non-ordinary matter ($-1\leq k<0$) the subcases (c1, c2) are impossible to hold. Thus, non-ordinary fluids cannot have relativistic flow at spatial infinity.

In the following we will analyze the behavior of the fluid by taking different cases for the state parameter $k$. For instance, we have $k=1$ (ultra-stiff fluid), $k=1/2$ (ultra-relativistic fluid), $k=1/3$ (radiation fluid) and $k=1/4$ (sub-relativistic fluid). For the case of the metric~\eqref{1met}, Eq. (\ref{cp2}) reduces to
\begin{equation}\label{cp3a}
    k=\frac{(3 \beta -2 \Lambda  r_c)r_c^2+6 M}{3 [(4+5 \beta  r_c-2 \Lambda  r_c^2) r_c-6 M]},
\end{equation}
and we keep in mind that $a^2=k$ in~\eqref{nds}. The system~\eqref{nds} and~\eqref{cp3a} form our basic equations for the remaining part of this section, which is devoted to applications. We mainly focus on anti-de Sitter-like $\text{f}(R)$ black holes with an application to Schwarzschild black hole. Further applications to anti-de Sitter-like and de Sitter-like $\text{f}(R)$ black holes with polytropic EOS for the test fluids are given in Sec.~\ref{secPTS}.

\subsection{Solution for ultra-stiff fluid (${\pmb k=1}$)}

Ultra-stiff fluids are those fluids in which isotropic pressure and energy density are equal.
For instance, the usual equation of state for the ultra-stiff fluids is $p = ke$ i.e. the value of state parameter is defined as $k=1$. This reduces~\eqref{cp2} or~\eqref{cp3a} to $f_c=0$, thus $r_c=r_h$~(\ref{3a},\ref{3b}). The Hamiltonian~\eqref{nds} reduces to
\begin{equation}\label{k11}
\mathcal{H}=\frac{1}{v^2r^4}.
\end{equation}
Since the Hamiltonian in Eq.(\ref{k11}) is a constant, one immediately obtains\footnote{For the cases $k=1$ and $k=1/2$ we have expressed explicitly $v$ as a function of $r$ as in Eqs.~\eqref{k11n} and~\eqref{k12v2}; it is possible to do the same for the other cases $k=1/3$ and $k=1/4$ [see Eqs.~\eqref{k131} and~\eqref{k141}] but the expressions of $v(r)$ would be cumbersome. That's why we preferred a numerical analysis in this section. It is worth mentioning that the expressions~\eqref{k11n} and~\eqref{k12v2} may be derived from the metric and the conservation laws using the classical approach for accretion~\cite{5u}.}
\begin{equation}\label{k11n}
    v\sim 1/r^{2}.
\end{equation}

It is clear from~\eqref{k11} that the point $(r,v^2)=(r_h,1)$ is not a CP of the dynamical system, as was noticed in the previous section. Notice that $\mathcal{H}$ no longer depends on $f$; thus, this expression, and the following conclusions, are valid for any metric of the form~\eqref{1}.
\begin{figure}[!htb]
\centering
\includegraphics[width=0.47\textwidth]{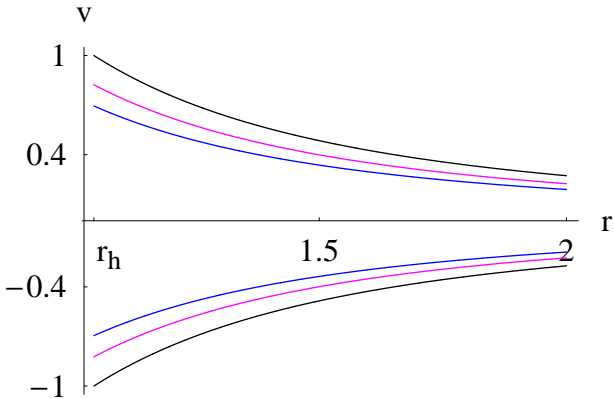}\\
\caption{{\footnotesize Contour plot of $\mathcal{H}$~\eqref{k11}, which is the simplified expression of $\mathcal{H}$~\eqref{nds}, for an anti-de Sitter-like $\text{f}(R)$ black hole $k=1$, $M=1$, $\bt =0.85$, $\La =-0.075$. The parameters are $r_h\simeq 1.04439$. Black plot: the solution curve through the CP for which $\mathcal{H}=\mathcal{H}_{\text{min}}=r_h^{-4}\simeq 0.84053$. Magenta plot: the solution curve for which $\mathcal{H}=\mathcal{H}_{\text{min}} + 0.4$. Blue plot: the solution curve for which $\mathcal{H}=\mathcal{H}_{\text{min}} + 0.9$.}}\label{Fig1}
\end{figure}

From~\eqref{k11} we see that, for physical flows ($|v|<1$), the lower value of $\mathcal{H}$ is $\mathcal{H}_{\text{min}}=1/r_h^4$: $\mathcal{H}>\mathcal{H}_{\text{min}}$. As shown in Fig.~\ref{Fig1}, physical flows are represented by the curves sandwiched by the two black curves, which are contour plots of $\mathcal{H}(r,v)=\mathcal{H}_{\text{min}}$. The upper curves where $v>0$ correspond to fluid outflow or particle emission and the lower curves where $v<0$ correspond to fluid accretion.

If $\mathcal{H}_0>\mathcal{H}_{\text{min}}$ is the value of the Hamiltonian on a solution curve, then in the ($r,v$) plane the curve is the plot $v=\pm 1/(\sqrt{\mathcal{H}_0}r^2)$. Using this we can evaluate all the other quantities, for instance~\eqref{b10} becomes
\begin{equation}\label{k12}
\Big(\frac{n}{n_c}\Big)^{2}=\frac{r_h^5f_{,r}|_{r=r_h}}{4}~\frac{\mathcal{H}_0r^4-1}{r^4f},
\end{equation}
for any solution curve $\mathcal{H}_0>\mathcal{H}_{\text{min}}= r_h^{-4}$, and
\begin{equation}\label{k13}
\Big(\frac{n}{n_c}\Big)^{2}=\frac{r_cf_{c,r_c}}{4}~\frac{1-v^2}{f}=\frac{r_hf_{,r}|_{r=r_h}}{4}~\frac{r^4-r_h^4}{r^4f},
\end{equation}
for the solution curve through $(r,v^2)=(r_h,1)$ ($\mathcal{H}_0=\mathcal{H}_{\text{min}}$), which all depend on $f$.

A contour plot of $\mathcal{H}$~\eqref{k11}, depicted in Fig.~\ref{Fig1}, shows two type of motion: (a) purely subsonic accretion (black, magenta, or blue curves where $v<0$) or subsonic flowout (black, magenta, or blue curves where $v>0$) for $\mathcal{H}>\mathcal{H}_{\text{min}}=r_h^{-4}$, and (b) purely supersonic accretion or flowout (along the red and green curves) for $\mathcal{H}<\mathcal{H}_{\text{min}}=r_h^{-4}$. The flow in (b), along the green and red curves, is however unphysical, for the speed of the flow exceeds that of light on some portions of the curves.  A brief elaboration is given in Table \ref{table2}.
\begin{table}
{\footnotesize
	\begin{center}
		\begin{tabular}{|c|l|}
			\hline
			\bf{Types} & \bf{Flow behavior}\\
			\hline
			I  & $\mathcal{H}>\mathcal{H}_{\text{min}}=r_{h}^{-4}$: Subsonic flow for $v<0$ and $v>0$ \\
			\hline
			II & $\mathcal{H}<\mathcal{H}_{\text{min}}=r_{h}^{-4}$: Unphysical flow \\
			\hline
		\end{tabular}
	\end{center}
}
\caption{{\footnotesize Types of flow on a solution curve for $k=1$ (Fig.~\ref{Fig1}).}}\label{table2}
\end{table}

\subsection{Solution for ultra-relativistic fluid (${\pmb k=1/2}$)}

Ultra-relativistic fluids are those fluids whose isotropic pressure is less than the energy density.
In this case, the equation of state is defined as $p=\frac{e}{2}$ yielding $k=1/2$.
Using this expression in~\eqref{cp3a} reduces to
\begin{eqnarray}\label{38}
Q(r_c)=\frac{\Lambda}{6}r_c^{3}-\frac{3\beta}{4}r_c^{2}-r_c+\frac{5}{2}M=0.
\end{eqnarray}
This polynomial has always one and only one positive root if $\La<0$ and $\bt\geq 0$. Converting this polynomial into the Weierstrass one $w(z)$ by the transformation $r_c=z+3\bt/(2\La)$, the CP $r_c$ is given either by (see Appendix A)
\begin{equation}\label{3aa}
r_c=\frac{3\bt}{2\La}+\sqrt{\frac{g_2}{3}}\cos\Big(\frac{\eta}{3}\Big),
\end{equation}
if $Q=0$ has at least two real roots or by
\begin{multline}\label{3ba}
r_c=\frac{3\bt}{2\La}+\frac{1}{2\cdot 9^{1/3}}[(9 g_3+ \sqrt{3} \sqrt{-\Delta })^{1/3}\\+(9 g_3-\sqrt{3} \sqrt{-\Delta })^{1/3}],
\end{multline}
if $Q=0$ has only one real root. Here $g_2$ and $g_3$ are defined by
\begin{equation*}
g_2=\frac{3 (9 \beta ^2+8 \Lambda )}{\Lambda ^2},\quad g_3=\frac{27 \beta ^3+36 \beta  \Lambda -60 M \Lambda ^2}{\Lambda ^3},
\end{equation*}
and $\De$ and the angle $0\leq \eta\leq \pi$ are defined as in Eqs.~\eqref{df2} and~\eqref{3.4}, respectively.
\begin{figure}[!htb]
\centering
\includegraphics[width=0.47\textwidth]{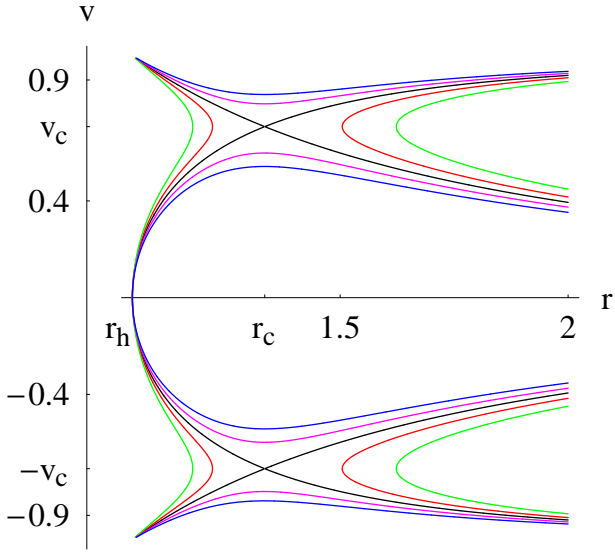}\\
\caption{{\footnotesize Contour plot of $\mathcal{H}$~\eqref{nds} for an anti-de Sitter-like $\text{f}(R)$ black hole $k=1/2$, $M=1$, $\bt =0.85$, $\La =-0.075$. The parameters are $r_h\simeq 1.04439$, $r_c\simeq 1.33467$, $v_c=1/\sqrt{2}\simeq 0.707107$. Black plot: the solution curve through the saddle CPs $(r_{c},v_c)$ and $(r_{c},-v_c)$ for which $\mathcal{H}=\mathcal{H}_c\simeq 0.926185$. Red plot: the solution curve for which $\mathcal{H}=\mathcal{H}_c - 0.04$. Green plot: the solution curve for which $\mathcal{H}=\mathcal{H}_c - 0.09$. Magenta plot: the solution curve for which $\mathcal{H}=\mathcal{H}_c + 0.04$. Blue plot: the solution curve for which $\mathcal{H}=\mathcal{H}_c + 0.09$.}}\label{Fig2}
\end{figure}

In the limit $\beta\rightarrow 0$, we recover the Schwarzschild anti-de Sitter spacetime and Eq.~(\ref{3aa}) reduces to
\begin{eqnarray}\label{z1}
r_{c}&=&\sqrt{\frac{g_2}{3}}\cos\Big(\frac{\eta}{3}\Big).
\end{eqnarray}

The Hamiltonian~\eqref{nds} takes the simple form
\begin{equation}\label{k121}
\mathcal{H}=\frac{\sqrt{f}}{r^2|v|\sqrt{1-v^2}}.
\end{equation}
It is clear from this expression that the point $(r,v^2)=(r_h,1)$ is not a CP of the dynamical system.
For some given value of $\mathcal{H}=\mathcal{H}_0$, Eq.~\eqref{k121} can be solved for $v^2$. We find
\begin{equation}\label{k12v2}
   v^2=\frac{1\pm \sqrt{1-4g(r)}}{2},
\end{equation}
where $g(r)\equiv f/(\mathcal{H}_0r^4)$. The plot in Fig.~\ref{Fig2} depicts, instead, $v$ versus $r$ for $M=1$, $\bt =0.85$, and $\La =-0.075$ resulting in $r_c\simeq 1.33467$ and $\mathcal{H}_c\simeq 0.926185$. The five solution curves, shown in Fig.~\ref{Fig2}, correspond to $\mathcal{H}_0=\{\mathcal{H}_c,\mathcal{H}_c\pm 0.04,\mathcal{H}_c\pm 0.09\}$. The upper plot for $v>0$ corresponds to fluid outflow or particle emission and that for $v>0$ corresponds to fluid accretion. The plot shows four types of fluid motion. (1) purely supersonic accretion ($v<-v_c$), which ends inside the horizon, or purely supersonic outflow ($v>v_c$); (2) purely subsonic accretion followed by subsonic flowout, this is the case of the branches of the blue and magenta solution curves corresponding to $-v_c<v<v_c$. Notice that for this motion the fluid reaches the horizon, $f(r_h)=0$, with vanishing speed ensuring that the Hamiltonian~\eqref{k121} remains constant. The critical black solution curve reveals two types of motions: if we assume that $dv/dr$ is continuous at the CPs, then (3) we have a supersonic accretion until ($r_c,-v_c$), followed by a subsonic accretion until ($r_h,0$), where the speed vanishes, then a subsonic flowout until ($r_c,v_c$), followed by a supersonic flowout, or (4) (lower plot) a subsonic accretion followed by a supersonic accretion which ends inside the horizon. In the upper plot, we have a supersonic outflow followed by a subsonic motion. The summary of this is given in Table \ref{table1}.
\begin{table*}
{\footnotesize
	\begin{center}
		\begin{tabular}{|c|l|}
			\hline
			\bf{Types} & \bf{Flow behavior}\\
			\hline
			I  & Supersonic for $-1<v<v_{c}$ and $1>v>v_{c}$ \\
			\hline
			II & Subsonic for $-v_{c}<v<v_{c}$ \\
			\hline
			III & Critical supersonic accretion until $(r_{c},-v_{c})$, subsonic flow from $(r_{c},-v_{c})$ until
                  $(r_{c},v_{c})$, suppersonic flowout \\
			\hline
			IV & Subsonic accretion until $(r_{c},-v_{c})$ then supersonic \\
			\hline
			V & Supersonic flowout until $(r_{c},v_{c})$ then subsonic \\
			\hline
		\end{tabular}
	\end{center}
}
\caption{{\footnotesize Different behaviors of the fluid flow for $k=1/2$ (Fig.~\ref{Fig2}).}}\label{table1}
\end{table*}

The fluid flow in Type (3) from ($r_c,-v_c$) to ($r_c,v_c$) describes a heteroclinic orbit that passes through two different saddle CPs: ($r_c,-v_c$) and ($r_c,v_c$). It is easy to show that the solution curve from ($r_c,-v_c$) to ($r_c,v_c$) reaches ($r_c,v_c$) as $\bar{t}\to -\infty$, and the curve from ($r_c,v_c$) to ($r_c,-v_c$) reaches ($r_c,-v_c$) as $\bar{t}\to +\infty$; we can change the signs of these two limits upon performing the transformation $\bar{t}\to -\bar{t}$ and $\mathcal{H}\to -\mathcal{H}$.

The flowout of the fluid, which starts at the horizon, is caused by the high pressure of the fluid, which diverges there~\eqref{pr2}: The fluid under effects of its own pressure flows back to spatial infinity.

\begin{figure}[!htb]
\centering
\includegraphics[width=0.47\textwidth]{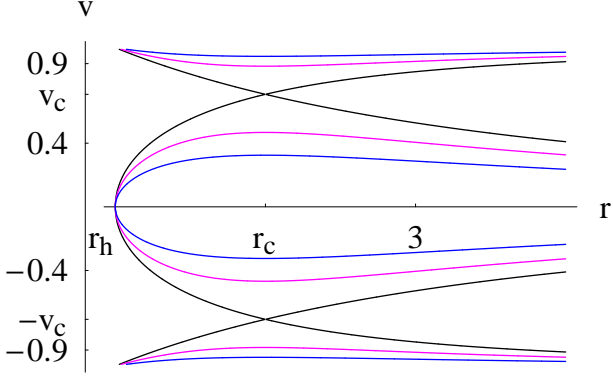}\\
\caption{{\footnotesize Contour plot of $\mathcal{H}$~\eqref{nds} for a Schwarzschild black hole with $k=1/2$, $M=1$, $\bt =0$, $\La =0$. The parameters are $r_h\simeq 2$, $r_c\simeq 2.5$, $v_c=1/\sqrt{2}\simeq 0.707107$. Black plot: the solution curve through the saddle CPs $(r_{c},v_c)$ and $(r_{c},-v_c)$ for which $\mathcal{H}=\mathcal{H}_c\simeq 0.143108$. Magenta plot: the solution curve for which $\mathcal{H}=\mathcal{H}_c + 0.03$. Blue plot: the solution curve for which $\mathcal{H}=\mathcal{H}_c + 0.09$.}}\label{Fig3}
\end{figure}

It is clear from Fig.~\ref{Fig2} that, after watching the subsonic branches of the blue and magenta solution curves, there is no way to support the claim, recalled at the end of Sec.~\ref{secHS}, that ``\texttt{the flow must be supersonic at the horizon}"~\cite{Sandip}. For these new solutions the speed of the fluid increases during the accretion from 0, according to the analysis made from~\eqref{gf1} to~\eqref{gf7b}, to some value below $v_c$ where $dv/dr=0$, then decreases to 0 at the horizon, and the process is reversed during the flowout. It is easy to show, using~\eqref{k12v2}, that the point where the speed is maximum is $r_c$, as shown in Fig.~\ref{Fig2}. Thus, the flow does not necessary become supersonic nor transonic near the horizon~\cite{NT1,NT2}. This conclusion does not depend on the presence of a negative cosmological nor on a nonvanishing constant $\bt$: such solutions exist even for a Schwarzschild black hole, as the subsonic branches of the blue and magenta solution curves in Fig.~\ref{Fig3} show.

Curiously enough, such solutions were never discussed in the literature. This is probably due to the fact that the pioneering works on this subject did not employ the Hamiltonian dynamical system approach to tackle the problem. These new solutions are related to the instability and fine tuning problems in dynamical systems. To see that consider the asymptotic behavior of~\eqref{k121}. Since $f\sim -(\La/3)r^2$ as $r\to\infty$ and since $\mathcal{H}$ remains constant on a solution curve, we must have $v\sim v_1r^{-1}$ ($v_1<0$ during accretion), which agrees with~\eqref{gf1} and~\eqref{gf5a}. Asymptotically, Eq.~\eqref{k121} reads
\begin{equation}\label{k123}
   \mathcal{H}\sim \mathcal{H}_{\infty}\equiv\frac{\sqrt{-\La/3}}{|v_1|},
\end{equation}
which is used to determine the value of $|v_1|$ by
\begin{equation}\label{k124}
   |v_1|=\frac{\sqrt{-\La/3}}{\mathcal{H}_{\infty}}.
\end{equation}
Notice that as $|v_1|$ increases, $\mathcal{H}_{\infty}$ decreases. Now consider the lower plot of Fig.~\ref{Fig2} and the branch of the black critical curve where first the speed is subsonic until the CP then it becomes supersonic. On this curve $\mathcal{H}\sim \mathcal{H}_{\infty}=\mathcal{H}_c$, it follows that
\begin{equation}\label{k125}
   |v_{1\text{b}}|=\frac{\sqrt{-\La/3}}{\mathcal{H}_c},
\end{equation}
where the subscript ``b" is for black. If one decreases the value of the asymptotic speed, that is, the value of $|v_1|$ by $\ep$: $|v_1|\to |v_{1\text{b}}|-\ep$, as is the case of the subsonic magenta curve of Fig.~\ref{Fig2}, then $\mathcal{H}_{\infty}$ increases by a corresponding amount: $\mathcal{H}_{\infty}\to \mathcal{H}_c+\ep\sqrt{-\La/3}/|v_{1\text{b}}|^2$. This small perturbation in the value of $|v_1|$ leads the flow to completely change course, by deviating from the black critical curve, and to undergo a purely subsonic motion along the subsonic magenta curve. Conversely, a small increase in the value of the asymptotic speed (of the coefficient $|v_1|$) would lead the flow to follow the red curve adjacent to the black critical curve. Thus, the black critical curve is certainly unstable and in practical situations it would not be easy to fix the value of $|v_1|$, which is an average value for the pressure is not zero, by fine tuning it to have a critical motion, that is, a motion that becomes supersonic beyond the CP and reaches the speed of light as the fluid approaches the horizon.

This stability issue is related to the character of the CPs ($r_c,-v_c$) and ($r_c,v_c$) that are saddle points of the Hamiltonian function. As is well known saddle points of the Hamiltonian function are also saddle points of the Hamiltonian dynamical system. Further analysis of stability requires linearization of the dynamical system and/or use of Lyapunov's theorems~\cite{book1,book2,book3} and their variants~\cite{variant}.

Another type of instability is the flowout that starts in the vicinity of the horizon ($r=r_h+0^+,v=0^+$) under the effect of a divergent pressure. This flowout is unstable, for it may follow a subsonic path (the magenta or blue curves) or a critical path (the black curve) through the CP ($r_c,v_c$) and becomes supersonic with a speed approaching that of light. From a cosmological point of view, this point ($r=r_h,v=0$) looks like an attractor where solution curves converge and a repeller from where the curves diverge~\cite{variant}.

The motion along the rightmost branches of the green and red curves is unphysical. Along the leftmost branches of these curves, we have an accretion starting from the leftmost point of the branch until the horizon where the speed vanishes and the pressure diverges, followed by a flowout back to the same starting point. To realize such a flow one needs to have a sink and source at the leftmost point of these branches.

\subsection{Solution for radiation fluid (${\pmb k=1/3}$)}

Radiation fluid is the fluid which absorbs the radiation emitted by the black hole. It is the most interesting case in astrophysics. Here, the value of state parameter $k=1/3$. Eq.~\eqref{cp3a} leads us to
\begin{equation}\label{39a}
\bt r_c^{2}+2r-6M=0,
\end{equation}
which is solved by
\begin{equation}\label{39b}
r_c=\frac{\sqrt{1+6\bt M}-1}{\bt}.
\end{equation}

The Hamiltonian~\eqref{nds} takes the simple form
\begin{equation}\label{k131}
\mathcal{H}=\frac{f^{2/3}}{r^{4/3}|v|^{2/3}(1-v^2)^{2/3}}.
\end{equation}
It is clear from this expression that the point $(r,v^2)=(r_h,1)$ is not a CP of the dynamical system. Eq.~\eqref{k131} can be solved for $v^2$, and a contour plot of it can be depicted, which reveals the same characteristics of the plot shown in Fig.~\ref{Fig2}; We observe the same types of motion as in the case $k=1/2$.

\subsection{Solution for sub-relativistic fluid (${\pmb k=1/4}$): Separatrix heteroclinic flows}

Sub-relativistic fluids are those fluids whose energy density exceeds their isotropic pressure. Taking the value of the state parameter $k=1/4$, Eq.~\eqref{cp3a} leads to
\begin{equation}\label{40}
N(r_c)=\Lambda r_c^{3}+\frac{3\beta}{2}r_c^{2}+6r_c-21M=0.
\end{equation}
This polynomial has either two distinct positive roots or a double positive root if $\La<0$ and $\bt\geq 0$. Converting this polynomial into the Weierstrass one $w(z)$ by the transformation $r_c=z-\bt/(2\La)$, the two CPs $r_{c1}<r_{c2}$ are given by (see Appendix A)
\begin{multline}\label{3aaa}
r_{c2}=\sqrt{\frac{g_2}{3}}\cos\Big(\frac{\eta}{3}\Big)-\frac{\bt}{2\La},\\ r_{c1}=-\sqrt{\frac{g_2}{3}}\cos\Big(\frac{\pi+\eta}{3}\Big)-\frac{\bt}{2\La},
\end{multline}
where $g_2$ and $g_3$ are defined by
\begin{equation*}
g_2=\frac{3 (\beta ^2-8 \Lambda )}{\Lambda ^2},\quad g_3=\frac{-\beta ^3+12 \beta  \Lambda +84 M \Lambda ^2}{\Lambda ^3},
\end{equation*}
and $\De$ and the angle $0\leq \eta\leq \pi$ are defined as in Eqs.~\eqref{df2} and~\eqref{3.4}, respectively.
\begin{figure}[!htb]
\centering
\includegraphics[width=0.47\textwidth]{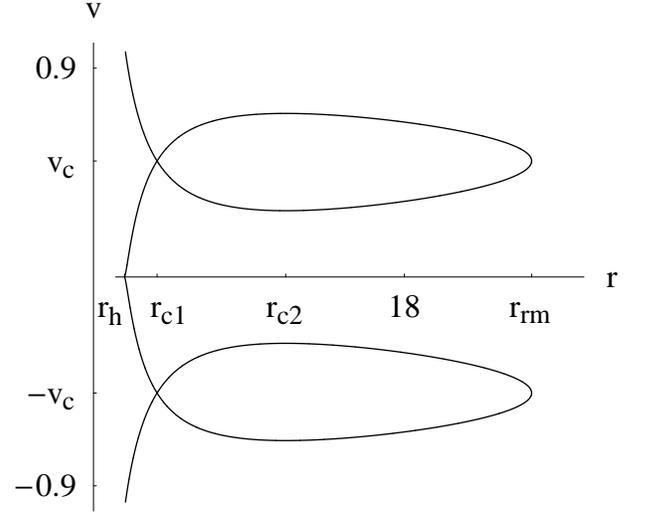}\\
\caption{{\footnotesize Contour plot of $\mathcal{H}$~\eqref{nds} for an anti-de Sitter-like $\text{f}(R)$ black hole with $k=1/4$, $M=1$, $\bt =0.05$, $\La =-0.04$. The parameters are $r_h\simeq 1.76955$, $r_{c1}\simeq 3.65928$, $r_{c2}\simeq 11.119$, $v_c=1/2$, $r_{\text{rm}}\simeq 25.3831$. The plot shows the heteroclinic solution curve through the saddle CPs $(r_{c1},v_c)$ and $(r_{c1},-v_c)$ for which $\mathcal{H}=\mathcal{H}(r_{c1},v_c)=\mathcal{H}(r_{c1},-v_c)\simeq 0.411311$. The two other CPs, $(r_{c2},v_c)$ and $(r_{c2},-v_c)$, are centers where $\mathcal{H}=\mathcal{H}(r_{c2},v_c)=\mathcal{H}(r_{c2},-v_c)\simeq 0.411311$.}}\label{Fig4}
\end{figure}

The Hamiltonian~\eqref{nds} takes the simple for
\begin{equation}\label{k141}
\mathcal{H}=\frac{f^{3/4}}{r\sqrt{|v|}(1-v^2)^{3/4}}.
\end{equation}
It is clear from this expression that the point $(r,v^2)=(r_h,1)$ is not a CP of the dynamical system. A contour plot of $\mathcal{H}$~\eqref{k141} is depicted in Fig.~\ref{Fig4} in the ($r,v$) plane. There are two saddle points $(r_{c1},v_c)$ and $(r_{c1},-v_c)$ and two centers $(r_{c2},v_c)$ and $(r_{c2},-v_c)$. Let $(r_{\text{rm}},v_c)$ and $(r_{\text{rm}},-v_c)$ be the rightmost points of the upper and lower plots, respectively. If we assume that $dv/dr$ remains continuous as the fluid crosses the saddle CPs, the accretion motion starts from the rightmost point $(r_{\text{rm}},-v_c)$ on the black curve in the lower plot. If the motion is subsonic it proceeds along the upper branch in the lower plot, goes through the CP $(r_{c1},-v_c)$, then crosses the horizon.

Otherwise, if the motion is supersonic it proceeds along the lower branch in the lower plot, goes again through the CP $(r_{c1},-v_c)$ until $v$ vanishes as the fluid approaches the horizon [this is obvious from~\eqref{k141} where $v$ vanishes whenever $f$ does too], then the fluid goes again through the CP $(r_{c1},v_c)$ and follows the upper branch of the upper plot undergoing a supersonic motion until the rightmost point of the upper plot $(r_{\text{rm}},v_c)$. First, by similar arguments as those given in the case $k=1/2$, it can be shown that such motion is unstable. Secondly, the motion may become periodic but it is too hard to achieve that by (a) fine tuning the speed of the fluid at $(r_{\text{rm}},-v_c)$ and (b) realizing a source at $(r_{\text{rm}},-v_c)$ and a sink at $(r_{\text{rm}},v_c)$.

The fluid flow along the branch of the curve from ($r_c,-v_c$) to ($r_c,v_c$) describes a heteroclinic orbit that passes through two different saddle CPs: ($r_c,-v_c$) and ($r_c,v_c$). It is easy to show that as the flow approaches, from within the heteroclinic orbit, one or the other saddle CP the dynamical-system's time $\bar{t}$ goes to $\pm\infty$.

Here again the flowout of the fluid, which starts at the horizon, is caused by the high pressure of the fluid, which diverges there~\eqref{pr2}.

As we have done in the case $k=1/2$, we consider the fluid flow where $r$ decreases but $v>0$ or $r$ increases but $v<0$ as unphysical since the fluid is taken as a test matter and we have neglected its backreaction on the metric of the black hole. As far as a fluid element is taken as a test particle, such a motion is not possible in the background of the black hole metric. This is why a flow along a closed path in Fig.~\ref{Fig4}, or ``homoclinic" as some authors call it, is unphysical. We do not know if homoclinic orbits exist in a more realistic model where the backreaction of the fluid is taken into consideration.

For the clarity of the plot, Fig.~\ref{Fig4} has been plotted for unphysical parameters $M=1$, $\bt =0.5$, and $\La =-0.075$; for astrophysical values of the the parameters ($\La\to 0^-$), the difference $r_{c2}-r_{c1}$ becomes so large to be represented on a sheet of paper. The constraint that two CPs exist is to have two positive roots for the polynomial in~\eqref{40}: $N(r)=\Lambda r^{3}+\frac{3\beta}{2}r^{2}+6r-21M$. With $\La <0$ and $\bt>0$, the polynomial has a local minimum (at some negative value of $r$) and a local maximum at
\begin{equation}\label{max1}
    r_s=-\frac{\sqrt{\bt^2-8\La}+\bt}{2\La}.
\end{equation}
The heteroclinic orbit exists if $N(r_c)=0$ has two positive CPs; that is, if $N(r_s)>0$ yielding
\begin{equation}\label{max2}
M<\frac{(\bt^2-8\La)^{3/2}+\bt^3-12\bt\La}{84\La^2},
\end{equation}
generalizing the expression derived in Ref.~\cite{t2}. This should be read as a constraint on $\bt$. In the limit $\La\to 0^-$, this reduces to
\begin{equation}\label{max3}
    \bt^3 >42M\La^2,
\end{equation}
and the expressions of the two positive CPs and the horizon read
\begin{multline}\label{max4}
   r_{c1}\simeq \frac{\sqrt{4+14M\bt}-2}{\bt},\quad r_{c2}\simeq -\frac{3\bt}{2\La},\\
 r_h\simeq \frac{\sqrt{1+8M\bt}-1}{2\bt}.\qquad\qquad\qquad\qquad\qquad\quad
\end{multline}
It is easy to show that $r_{c1}>r_h$.

In the astrophysical limit $\La\to 0^-$ we find, for general values of $k$, the following constraints on $\bt$
\begin{equation}\label{max5}
\left\{
  \begin{array}{ll}
   \bt >\frac{42M\La^2(1-3k)^3}{(1-5k)^2(5-19k)} & \;\hbox{$\frac{1}{5}<k<\frac{5}{19}$;} \\
    \bt >\frac{2\sqrt{-\La}}{3}~(21M\sqrt{-\La}-5) & \;\hbox{$k=\frac{1}{5}$.}
  \end{array}
\right.
\end{equation}
In this limit, the CPs are expressed as
\begin{equation}\label{max6}
r_{c1}\simeq \left\{
  \begin{array}{ll}
   \frac{\sqrt{k^2 (4+30 M \beta )+4 k M \beta -2 M \beta }-2 k}{(5 k-1) \beta } & \;\hbox{$\frac{1}{5}<k<\frac{5}{19}$;} \\
    4M(1-16M^2 \La/3) & \;\hbox{$k=\frac{1}{5}$,}
  \end{array}
\right.
\end{equation}
\begin{equation}\label{max7}
r_{c2}\simeq \left\{
  \begin{array}{ll}
   \frac{3(1-5k)\bt}{2(1-3k)\La} & \;\hbox{$\frac{1}{5}<k<\frac{5}{19}$;} \\
    \frac{\sqrt{3}}{\sqrt{-\La}}& \;\hbox{$k=\frac{1}{5}$,}
  \end{array}
\right.
\end{equation}
while the expression of $r_h$~\eqref{max4} is independent of $k$.

\section{Polytropic test fluids\label{secPTS}}

A very interesting approach to describe the motion of fluid is by constructing its models.
The prototype of such model is \emph{Chaplygin gas}. The Chaplygin gas model leads to very interesting results. Some of them are discussed in Ref \cite{MCG1,MCG2,MCG3,MCG4,MCG5}.
There are many variations of the Chaplygin gas model have been proposed in the literature. One of them is the modified Chaplygin gas model~\cite{mod1,mod2}. In astrophysics, the modified Chaplygin gas is the most general exotic fluid. Its equation of state is:
\begin{eqnarray}\label{42}
p=A n-\frac{B}{n^{\alpha}},
\end{eqnarray}
where $A$ and $B$ are constants and $(0<\alpha<1)$. If we put $A=0$, $B=-k$ and $\alpha=-\gamma$, we get the polytropic equation of state i.e. $p=G(n)=\mathcal{K}n^{\gamma}$, where $\mathcal{K}$ and $\gamma$ are constants. For ordinary matter, one generally works with the constraint $\ga>1$. In this work, we only observe the constraint $\ga\neq 1$.

Inserting $p=G(n)=\mathcal{K}n^{\gamma}$ in the differential equation~\eqref{b5} yields
\begin{equation*}
   nF'-F=\mathcal{K}n^{\ga}.
\end{equation*}
The solution provides the energy density $e=F$ by
\begin{equation}\label{pl1}
  e=F(n)=mn+\frac{\mathcal{K}n^{\ga}}{\ga-1},
\end{equation}
where a constant of integration has been identified with the baryonic mass $m$. This yields~\eqref{b2}
\begin{equation}\label{pl2}
  h=m+\frac{\mathcal{K}\ga n^{\ga-1}}{\ga-1}.
\end{equation}
The three-dimensional speed of sound is found from~\eqref{19b} by
\begin{equation}\label{pl3}
   a^2=\frac{(\ga -1)X}{m(\ga -1)+X}\qquad (X\equiv \mathcal{K}\ga n^{\ga -1}).
\end{equation}
On comparing~\eqref{pl2} and~\eqref{pl3} we see that
\begin{equation}\label{pl4}
    h=m~\frac{\ga -1}{\ga -1-a^2},
\end{equation}
similar to an expression for $h$ derived for the accretion onto a black hole in a string cloud background~\cite{string}.

Using~\eqref{b10} in~\eqref{pl2}, we obtain
\begin{equation}\label{pl5}
    h=m\Big[1+Y\Big(\frac{1-v^2}{r^4fv^2}\Big)^{(\ga-1)/2}\Big],
\end{equation}
where
\begin{equation}\label{pl6}
   Y\equiv \frac{\mathcal{K}\ga n_c^{\ga-1}}{m(\ga-1)}~\Big(\frac{r_c^5f_{c,r_c}}{4}\Big)^{(\ga-1)/2}=\text{ const.}.
\end{equation}
Inserting~\eqref{pl5} into~\eqref{h3} we evaluate the Hamiltonian by
\begin{equation}\label{p17}
   \mathcal{H}=\frac{f}{1-v^2}~\Big[1+Y\Big(\frac{1-v^2}{r^4fv^2}\Big)^{(\ga-1)/2}\Big]^2,
\end{equation}
where $m^2$ has been absorbed into a re-definition of ($\bar{t},\mathcal{H}$).

A couple of remarks concerning the fluid flow onto an anti-de Sitter-like $\text{f}(R)$ black hole are in order. For ordinary matter $\mathcal{K}>0$ and $f_{c,r_c}>0$ (since  we are interested in the cases where $r_c>r_h$), this implies (a) $Y>0$ if $\ga>1$ or (b) $Y<0$ if $\ga<1$ ($\ga\neq0$).

For the case (a) the sum of the terms inside the square parentheses in~\eqref{p17} is positive while the coefficient $f/(1-v^2)$ diverges as $r\to\infty$ ($0\leq 1-v^2<1$). So, the Hamiltonian too diverges. Since the latter has to remain constant on a solution curve, we conclude that there are no global solutions in this case (solutions that extend to spatial infinity). This conclusion remains true even if $\La =0$ provided $\bt\neq 0$. If $\La =0$ and $\bt = 0$ (the Schwarzschild metric), the global solutions do not exist if $|v_{\infty}|=1$~\eqref{gf1} and exist otherwise provided $0<\al\leq 2$ if $|v_{\infty}|=0$ or $0<\al$ if $0<|v_{\infty}|<1$.
\begin{figure*}[!htb]
\centering
\includegraphics[width=0.47\textwidth]{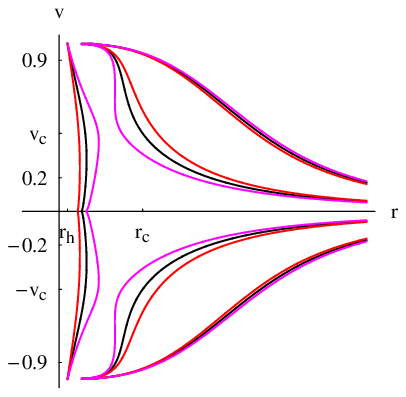} \includegraphics[width=0.47\textwidth]{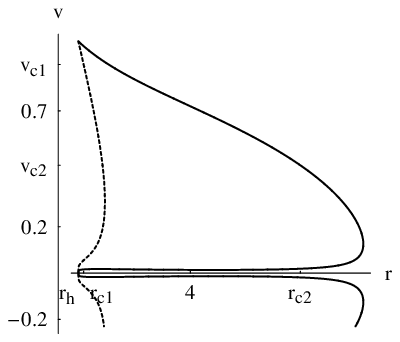}\\
\caption{{\footnotesize Left panel is a contour plot of $\mathcal{H}$~\eqref{p17} for an anti-de Sitter-like $\text{f}(R)$ black hole with $M=1$, $\bt =0.05$, $\La =-0.04$, $\ga=1/2$, $Y=-1/8$, $n_c=0.1$. The parameters are $r_h\simeq 1.76955$, $r_c\simeq 5.37849$, $v_c\simeq 0.464567$. Black plot: the solution curve through the CPs $(r_{c},v_c)$ and $(r_{c},-v_c)$ for which $\mathcal{H}=\mathcal{H}_c\simeq 0.379668$. Red plot: the solution curve for which $\mathcal{H}=\mathcal{H}_c - 0.09$. Magenta plot: the solution curve for which $\mathcal{H}=\mathcal{H}_c + 0.09$. Right panel is a contour plot of $\mathcal{H}$~\eqref{p17} for an anti-de Sitter-like $\text{f}(R)$ black hole with $M=1$, $\bt =0.05$, $\La =-0.04$, $\ga =5.5/3$, $Y=1/8$, $n_c=0.001$. The parameters are $r_h\simeq 1.76955$, $r_{c1}\simeq 1.87377$, $v_{c1}\simeq 0.900512$, $r_{c2}\simeq 6.19113$, $v_{c2}\simeq 0.465236$. Continuous black plot: the solution curve through the CPs $(r_{c2},v_{c2})$ and $(r_{c2},-v_{c2})$ for which $\mathcal{H}=\mathcal{H}_{c2}\simeq 1.94447$. Dashed black plot: the solution curve through the CPs $(r_{c1},v_{c1})$ and $(r_{c1},-v_{c1})$ for which $\mathcal{H}=\mathcal{H}_{c1}\simeq 0.443809$. For the clarity of the plot, we have partially removed the branches $v<0$.}}\label{Fig5}
\end{figure*}

For the case (b), since $Y<0$, we can make it such that
\begin{equation}\label{p18}
    1+Y\Big(\frac{1-v^2}{r^4fv^2}\Big)^{(\ga-1)/2}\propto r^{-1}\quad\text{ as }\quad r\to\infty,
\end{equation}
in order to have global solutions. For instance, if we restrict ourselves to $v$ having an expansion in powers of $1/r$ with a vanishing three-dimensional speed at spatial infinity~\eqref{gf1}
\begin{equation}\label{p19}
v\simeq v_1r^{-\al}+v_2r^{-\de}\ \text{ as }\ r\to\infty\qquad (\de>\al>0),
\end{equation}
then, on observing~\eqref{p18} we find $\al =3$, $\de\geq 4$, and
\begin{equation}\label{p20}
v_1^2=(-3/\La)(Y^2)^{1/(\ga-1)}.
\end{equation}
This is another, rather much harder, fine tuning problem. Here $Y$ depends on $n_c$, so is $v_1$: Unless $v_1^2$ is the rhs of~\eqref{p20}, there will be no global solutions to this case too.

For non-ordinary matter, since $\mathcal{K}<0$, the above two cases are reversed, that is, for $\ga>1$ it is possible to have global solutions, again with a fine tuning problem, while for $\ga<1$ ($\ga\neq 0$) there are non global solutions.

In the following we provide two curve solutions for an anti-de Sitter-like $\text{f}(R)$ black hole in the cases $\ga>1$ (non-global solution) and $\ga<1$ (global solution) and a curve solution for a de Sitter-like $\text{f}(R)$ black hole in the case $\ga>1$. First, using~\eqref{b10} we rewrite~\eqref{pl3} as
\begin{multline}\label{p21}
    \Big[\frac{n_c}{Y}~\Big(\frac{r_c^5f_{c,r_c}}{4}\Big)^{1/2}+\Big(\frac{1-v^2}{r^4fv^2}\Big)^{(\ga-1)/2}\Big]a^2\\
    =(\ga-1)\Big(\frac{1-v^2}{r^4fv^2}\Big)^{(\ga-1)/2}.
\end{multline}
Since at the CPs we have $a_c^2=v_c^2$~\eqref{cp1}, we replace $a^2$ in~\eqref{p21} and in~\eqref{cp2} by $v_c^2$ and solve the system~\eqref{p21} and~\eqref{cp2} to find the CPs ($r_c,v_c$). We rewrite these latter equations after making the substitution $a_c^2=v_c^2$ as
\begin{align}
\label{p22a}&(\ga-1-v_c^2)\Big(\frac{1-v_c^2}{r_c^4f_cv_c^2}\Big)^{(\ga-1)/2}
=\frac{n_c}{Y}\Big(\frac{r_c^5f_{c,r_c}}{4}\Big)^{1/2}v_c^2,\\
\label{p22b}&v_c^2=\frac{r_cf_{c,r_c}}{r_cf_{c,r_c}+4f_c}=\frac{(3 \beta -2 \Lambda  r_c)r_c^2+6 M}{3 [(4+5 \beta  r_c-2 \Lambda  r_c^2) r_c-6 M]}.
\end{align}
Here we keep using $f$ to show the general character of these equations. Inserting~\eqref{p22b} into~\eqref{p22a} we can first solve numerically for $r_c$ then get $v_c$ from~\eqref{p22b}. Since the signs of both sides of~\eqref{p22a} must be the same, we conclude that, for $\ga<1$, $v_c^2>\ga-1$ (which is always satisfied) and that, for $\ga>1$, $v_c^2<\ga-1$.
\begin{figure*}[!htb]
\centering
\includegraphics[width=0.47\textwidth]{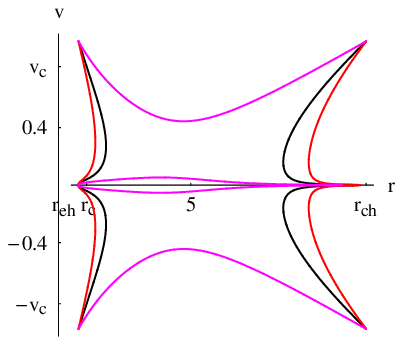} \includegraphics[width=0.47\textwidth]{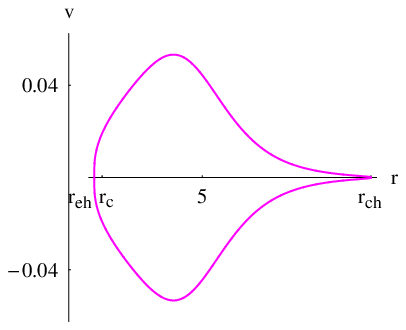}\\
\caption{{\footnotesize Left plot is a contour plot of $\mathcal{H}$~\eqref{p17} for a de Sitter-like $\text{f}(R)$ black hole with $M=1$, $\bt =0.05$, $\La =0.04$, $\ga =1.7$, $Y=1/8$, $n_c=0.001$. The parameters are $r_{eh}\simeq 1.91048$, $r_{ch}\simeq 9.8282$, $r_{c}\simeq 2.13406$, $v_{c}\simeq 0.824282$. Black plot: the solution curve through the CPs $(r_{c},v_{c})$ and $(r_{c},-v_{c})$ for which $\mathcal{H}=\mathcal{H}_{c}\simeq 0.390248$. Red plot: the solution curve corresponding to $\mathcal{H}=\mathcal{H}_{c}- 0.1$. Magenta plot: the solution curve corresponding to $\mathcal{H}=\mathcal{H}_{c}+ 0.29$. Right plot is a zoomed in plot of the cyclic flow corresponding to $\mathcal{H}=\mathcal{H}_{c}+ 0.29$.}}\label{Fig7}
\end{figure*}

Notice that the solution curves do not cross the $r$ axis at points where $v=0$ and $r\neq r_h$, for otherwise the Hamiltonian~\eqref{p17} would diverge there. We recall that $r_h$ is the unique horizon of an anti de Sitter-like $\text{f}(R)$ black hole or it represents either the event horizon $r_{eh}$ or the cosmological horizon $r_{ch}$ of a de Sitter-like $\text{f}(R)$ black hole. The curves may cross the $r$ axis at the unique point $r=r_h$ in the vicinity of which $v$ behaves as
\begin{multline}\label{p23}
    |v|\simeq |v_0||r-r_h|^{\frac{2-\ga}{2(\ga-1)}}\\
    \text{ with }\quad v_0^{2(\ga-1)}=\frac{Y^2f'(r_h)^{2-\ga}}{r_h^{4(\ga-1)}\mathcal{H}(r_h,0)},\quad
\end{multline}
if $f=0$ has a single root at $r_h$. We see that only solutions with $1<\ga<2$ may cross the $r$ axis. Here $\mathcal{H}(r_h,0)$ is the value of the Hamiltonian on the solution curve, which is the limit of $\mathcal{H}(r,v)$ as $(r,v)\to(r_h,0)$. This can be evaluated at any other point on the curve. The pressure $p=\mathcal{K}n^{\ga}$ diverges at the horizon as
\begin{equation}\label{p23b}
    p\propto |r-r_h|^{\frac{-\ga}{2(\ga-1)}}\qquad (1<\ga<2).
\end{equation}

For both plots of Fig.~\ref{Fig5} we took $M=1$, $\bt =0.05$, and $\La =-0.04$.

In the left panel of Fig.~\ref{Fig5}, we took $\ga=1/2$, $Y=-1/8$, and $n_c=0.1$, yielding one CP ($r_c\simeq 5.37849,v_c\simeq 0.464567$). We see from the graph that there are two types of fluid flow, an accretion which starts subsonic at spatial infinity and ends supersonic into the horizon (passing through the non-saddle CP or avoiding it), and a supersonic flowout from a neighborhood of the horizon which ends subsonic with gradually vanishing speed at spatial infinity according to (\ref{p19},\ref{p20}) (passing through the non-saddle CP or avoiding it). Along the leftmost branches we have an accretion starting from the leftmost point of the branch until the horizon where the speed vanishes and the pressure diverges, followed by a flowout back to the same starting point. Had we taken a lower number density $n_c=0.001$ we would still get the same types of flow but the uppermost, lowermost, and leftmost branches of the plot would disappear.

In the right panel of Fig.~\ref{Fig5}, we took $\ga=5.5/3$, $Y=1/8$, and $n_c=0.001$, yielding four CPs but none of them is a saddle point: ($r_{c1}\simeq 1.87377,v_{c1}\simeq 0.900512$), ($r_{c1},-v_{c1}$), ($r_{c2}\simeq 6.19113,v_{c2}\simeq 0.465236$), and ($r_{c2},-v_{c2}$). The right panel of Fig.~\ref{Fig5} shows a typical flow for these range of parameters ($\ga=5.5/3$, $Y=1/8$). There are three types of flow: subsonic non-global, non-relativistic (resp. more or less relativistic), and non-heteroclinic (for it does not pass through the CPs) accretion starting from the leftmost point of the continuous  (resp. dashed) branch until the horizon where the speed vanishes and the pressure diverges, followed by a non-relativistic (resp. more or less relativistic) flowout. This flow could be made periodic by realizing a source-sink at the rightmost point of the graph, as we have seen earlier. There are two other types of flow: partly subsonic and partly supersonic accretion and flowout along the continuous and dashed branches. The summary of this is given in Table \ref{table3}.
\begin{table}
{\footnotesize
	\begin{center}
		\begin{tabular}{|c|l|}
			\hline
			\bf{Types} & \bf{Flow behavior} \\
			\hline
			I  & Leftmost branches: Unphysical \\
			\hline
			II & Left panel: Critical transonic accretion and flowout\\
            \hline
			III  & Left panel: Non-critical sub-super sonic accretion and flowout \\
			\hline
			IV  & Right panel: Non-relativistic subsonic accretion and flowout \\
                & (with source-sink at the rightmost point of the graph)\\
			\hline
			V  & Right panel: Critical transonic accretion and flowout \\
               & (with source-sink at the rightmost point of the graph)\\
			\hline
		\end{tabular}
	\end{center}
}
\caption{{\footnotesize Behavior of flow for the polytropic equation of state in Fig \ref{Fig5}.}}\label{table3}
\end{table}
We emphasize that since the fluid is seen as a test matter in the geometry of the black hole, there is no homoclinic flow, that is, a flow following a closed curve in the right panel of Fig.~\ref{Fig5}.

In our next application we rather consider a de Sitter-like $\text{f}(R)$ black hole taking $M=1$, $\bt =0.05$, $\La =0.04$, $\ga =1.7$, $Y=1/8$, $n_c=0.001$ as in Fig.~\ref{Fig7}. For these values of the parameters, the dynamical system has two non-saddle CPs: $(r_{c}\simeq 2.13406,v_{c}\simeq 0.824282)$ and $(r_{c},-v_{c})$. The flow for $\mathcal{H}\leq \mathcal{H}_{c}\simeq 0.390248$ shows no difference than that of the right panel of Fig.~\ref{Fig5} corresponding to an anti-de Sitter-like $\text{f}(R)$ black hole. For $\mathcal{H}> \mathcal{H}_{c}$, we observe two types of flow connecting the two horizons, one of which is supersonic, relativistic, near the horizons and becomes subsonic midway of the horizons (uppermost and lowermost branches of the magenta curve). The other flow  connecting the two horizons is, rather, cyclic physical flow with vanishing speed at both the event $r_{eh}\simeq 1.91048$ and the cosmological $r_{ch}\simeq 9.8282$ horizons, as shown in the right plot of Fig.~\ref{Fig7}. There is no need to realize a source at one horizon and a sink at the other; this subsonic, non-relativistic, cyclic (non-homoclinic, for it does not pass through the CP) flow is maintained by the high, rather divergent~\eqref{p23b}, pressure at both horizons. If the fluid is hot, a two-temperature ion (plasma) would form and the cyclic flow becomes the source of energy radiation~\cite{BK}. If the fluid is multi-specie, each component would radiate at different frequency, resulting in a spectrum characteristic of the fluid composition. The higher the value of the Hamiltonian the lower is the speed of flow along the closed branch.

From our above formulas we can make a good estimate of the proper period and frequency of such a cyclic flow. Assuming $v^2\ll 1$, that is, a relatively higher value of the Hamiltonian, then~\eqref{p17} reduces to
\begin{equation}\label{p24}
   (v\sqrt{f})^{\ga-1}\simeq \frac{Y}{r^{2(\ga-1)}(\sqrt{\mathcal{H}_{\text{cyc}}/f}-1)},
\end{equation}
where $\mathcal{H}_{\text{cyc}}$ is the value of the Hamiltonian that generates the cyclic flow between the event and cosmological horizons. The first equation in~\eqref{v2}
\begin{equation}\label{p25}
  Y^{\frac{1}{\ga-1}}d\tau \simeq r^2 \Big(\sqrt{\mathcal{H}_{\text{cyc}}/f}-1\Big)^{\frac{1}{\ga-1}}dr.
\end{equation}
The integral of the rhs of~\eqref{p25}, with the limits being ($r_{eh},r_{ch}$), converges if $\ga>3/2$ (recall that we are assuming that each horizon ($r_{eh},r_{ch}$), being a single root of $f=0$, is non-extremal) and diverges as $\ln |r-r_h|$ if $\ga=3/2$. For the values of Fig.~\ref{Fig7}, $\mathcal{H}_{\text{cyc}}=\mathcal{H}_c+0.29\simeq 0.680248$, we find the proper period to be
\begin{equation*}
\tau \simeq 2Y^{\frac{1}{1-\ga}}\int_{r_{eh}}^{r_{ch}}r^2 \Big(\sqrt{\mathcal{H}_{\text{cyc}}/f}-1\Big)^{\frac{1}{\ga-1}}dr\simeq 26761.9.
\end{equation*}

\section{Hu-Sawicki and Starobinsky models of $\text{f}(R)$ gravity\label{sec2m}}
\begin{figure*}[!htb]
\centering
\includegraphics[width=0.47\textwidth]{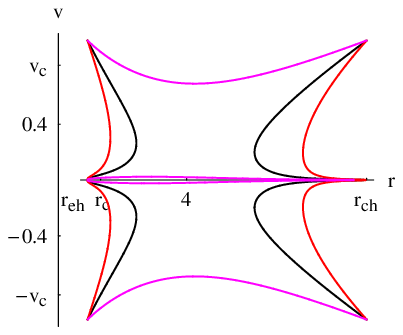} \includegraphics[width=0.47\textwidth]{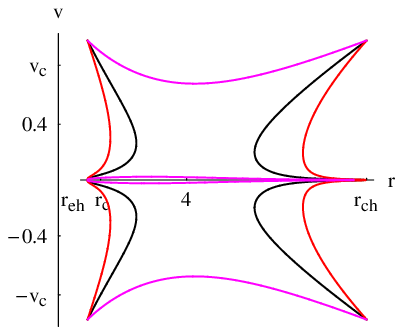}\\
\caption{{\footnotesize Contour plot of $\mathcal{H}$~\eqref{p17} for a de Sitter-like $\text{f}(R)$ black hole with $\text{f}(R)$ given by Hu-Sawicki formula~\eqref{hs5}. We took $M=1$, $Q =0.01$, $R_0 =0.16$, $\ga =1.7$, $Y=1/8$, $n_c=0.001$, $q_1=41$, $q_2=19$, and $c_1$ and $c_2$ are given by~\eqref{hs9}. Left plot: For $c_2$ we took the upper sign in~\eqref{hs9}, $\text{f}\,'(R_0) \simeq 1.96272$, $r_{eh}\simeq 2.12857$, $r_{ch}\simeq 7.39749$, $r_{c}\simeq 2.37452$, $v_{c}\simeq 0.822763$, and $\mathcal{H}=\mathcal{H}_{c}\simeq 0.291918$. Right plot: For $c_2$ we took the lower sign in~\eqref{hs9}, $\text{f}\,'(R_0) \simeq 0.0372803$, $r_{eh}\simeq 2.12854$, $r_{ch}\simeq 7.3975$, $r_{c}\simeq 2.37448$, $v_{c}\simeq 0.822764$, and $\mathcal{H}=\mathcal{H}_{c}\simeq 0.291918$.}}\label{Fig8}
\end{figure*}
$\text{f}\,'(R_0) =2R_0$,
Two more solution curves are provided in this section and concern two of the most popular models of $\text{f}(R)$ gravity: The Hu-Sawicki and Starobinsky models~\cite{m1,m2}.

There is a variety of black hole solutions of $\text{f}(R)$ gravity models, the most treated in the literature are constant curvature, $R=R_0$, solutions. If $R$ is the constant $R_0$, the field equations take the form
\begin{equation}\label{hs1}
    R_{\mu\nu}[1+\text{f}\,'(R_0)]-\tfrac{1}{2}g_{\mu\nu}[R_0+\text{f}\,(R_0)]=-8\pi T_{\mu\nu}.
\end{equation}
For an electromagnetic source,
\[
T^{\mu}_{\ \nu}=-\tfrac{1}{4\pi}\big(F^{\mu\al}F_{\nu\al}-\tfrac{1}{4}\de^{\mu}_{\ \nu}F^{\al\bt}F_{\al\bt}\big),
\]
(with $F_{\mu\nu}=\partial_{\mu}A_{\nu}-\partial_{\nu}A_{\mu}$) we have $T^{\mu}_{\ \mu}\equiv 0$. The trace of~\eqref{hs1} yields
\begin{equation}\label{hs2}
R_0+\text{f}\,(R_0)=[1+\text{f}\,'(R_0)]R_0/2,
\end{equation}
reducing~\eqref{hs1} to
\begin{equation}\label{hs3}
    \underbrace{R_{\mu\nu}-\tfrac{1}{2}R_0g_{\mu\nu}}_{G_{\mu\nu}}+\frac{R_0}{4}g_{\mu\nu}=
    -8\pi~\frac{T_{\mu\nu}}{1+\text{f}\,'(R_0)},
\end{equation}
where $G_{\mu\nu}$ is the Einstein tensor. On comparing~\eqref{hs3} with the field equations of general relativity, we see that $R_0/4$ plays the role of an effective cosmological constant and $T_{\mu\nu}/[1+\text{f}\,'(R_0)]$ is an effective SET. If the vector potential $A_{\mu}=(-Q/r,0,0,0)$, we obtain the spherically symmetric solution given by~\eqref{1} with\footnote{Equation~~\eqref{hs4} provides the correct expression of $f(r)$ of the solution given by Eq.~(32) of Ref.~\cite{m3}.}
\begin{equation}\label{hs4}
    f(r)=1-\frac{2M}{r}+\frac{Q^2}{[1+\text{f}\,'(R_0)]r^2}-\frac{R_0}{12}~r^2.
\end{equation}

\subsection{Starobinsky model}
This is the model with $\text{f}(R)=R^2/(6\mathcal{M}^2)$ where the constant $\mathcal{M}$ has value corresponding to the mass scale for quantum gravity. The only existing solution to~\eqref{hs2} is $R_0=0$ reducing~\eqref{hs4} to Reissner-Nordstr\"om black hole the fluid accretion onto which has already been investigated in the literature~\cite{Pacheco}, and is similar to the Schwarzschild case~\cite{5u}, so we won't comment on this case.

\subsection{Hu-Sawicki model}
This corresponds to
\begin{equation}\label{hs5}
\text{f}(R)=-\mathcal{M}^2\frac{c_1(R/\mathcal{M}^2)^n}{c_2(R/\mathcal{M}^2)^n+1},
\end{equation}
where $n>0$, ($c_1,c_2$) are proportional constants~\cite{m1}
\begin{equation}\label{hs6}
\frac{c_1}{c_2}\equiv q_2\approx 6~\frac{\Om_{\La}}{\Om_{m}}=6~\frac{0.76}{0.24}=19,
\end{equation}
and the mass scale
\begin{equation*}
\mathcal{M}^2=(8315\text{Mpc})^{-2}\Big(\frac{\Om_{m}h^2}{0.13}\Big).
\end{equation*}
At the present epoch~\cite{m1}
\begin{equation}\label{hs7}
\frac{R_0}{\mathcal{M}^2}\equiv q_1\approx\frac{12}{\Om_{m}}-9=41.
\end{equation}

For $n>0$, Eq.~\eqref{hs2} has always the root $R_0=0$. Notice that the model~\eqref{hs5} has been introduced in order to keep $|\text{f}\,'(R_0)|\ll 1$, which ensures stability. Hence, we rule out the case $0<n<1$ which would yield $|\text{f}\,'(R_0)|\to\infty$ as $R_0\to 0$. For $n\geq 1$, the root $R_0=0$ reduces~\eqref{hs4} to Reissner-Nordstr\"om black hole.

From now on we take $n=2$. Since we want that one of the other roots of~\eqref{hs2} be $R_0=q_1\mathcal{M}^2$, we substitute~\eqref{hs6} and~\eqref{hs7} into~\eqref{hs2} to obtain
\begin{equation}
q_1^3(q_1-2q_2)c_2^2+2q_1^2c_2+1=0,
\end{equation}
yielding
\begin{equation}\label{hs8}
c_1=q_2c_2,\qquad c_2=-\frac{1}{q_1^{3/2}(\sqrt{q_1}\pm\sqrt{2q_2})}.
\end{equation}
With the numerical values in~\eqref{hs6} and~\eqref{hs7}, the four values of $c_1$ and $c_2$ are all negative and one should keep those values that ensure $|\text{f}\,'(R_0)|\ll 1$
\begin{equation}\label{hs9}
c_1=q_2c_2,\qquad c_2=-\frac{1}{q_1^{3/2}(\sqrt{q_1}\pm\sqrt{2q_2})}.
\end{equation}

With $f(r)$ given by~\eqref{hs4}, the rhs of~\eqref{p22b} reads
\begin{equation}\label{hs10}
\hspace{-2mm}v_c^2=\frac{(1+\text{f}\,'(R_0)) (R_0 r_c^3-12 M) r_c+12 Q^2}{3 [(1+\text{f}\,'(R_0)) (R_0 r_c^3-8 r_c+12 M) r_c-4 Q^2]}.
\end{equation}

For the plots of Fig.~\ref{Fig8}, we used Eqs.~\eqref{p22a} and~\eqref{hs10} to find the critical points. The graphs show that accretion is insensitive to the values of the constants ($c_1,c_2$) and to the value of $\text{f}\,'(R_0)$ whose effect is to modify the value of the charge in~\eqref{hs4}.

\section{Conclusion}

We have developed a Hamiltonian dynamic system for tackling a variety of problems ranging from accretions, matter jets, particle emissions to cosmological and astrophysical applications whenever conservation laws apply. There are several choices for the dynamical variables arguments of the Hamiltonian. The advantage of using the three velocity is that this entity is bounded (by $-1$ and 1) and it does not diverge in contrast with the pressure and the baryon number density, and other densities, which may diverge on the horizons. Throughout the paper we kept using the metric coefficient $f(r)$ to emphasize the general character of the derived mathematical expressions. Since the scope of the model of accretion is fairly wide and applies to all static spherically symmetric solutions (asymptotically flat or else), the
present analysis can also be done for other $\text{f}(R)$ black holes as well as $\text{f}(T)$ black holes~\cite{FTG}. Due to the generality of our work, further analysis will be trivial.

Our general results that applies to all metrics of the form~\eqref{1} and to all perfect fluids, independently of the form of the EOS, are as follows. The Michel-type accretion of a perfect fluid is characterized by
\begin{itemize}
  \item The thermodynamic state functions are determined upon integrating a first order differential equation;
  \item If the three velocity vanishes on the horizon(s), the particle number density $n$ diverges there independently of the expression of $f$ and of that of the EOS. Since the specific enthalpy $h$ is never zero for ordinary matter, this implies that the sum $e+p$ diverges there at least as fast as $n$;
  \item The fluid may become ultra-stiff as it approaches the horizon(s).
\end{itemize}

By applying the Hamiltonian dynamic system to $\text{f}(R)$ gravity we have performed a detailed analysis of the Michel-type accretion onto a static spherically symmetric black hole in $\text{f}(R)$ gravity. Not every model of $\text{f}(R)$ theory can predict black holes unless the function $\text{f}(R)$ satisfies certain viability conditions such as $\text{f}\,'(R)>0$ and $\text{f}\,''(R)>0$, and asymptotically de Sitter phase at present time (see further details in~\cite{Amen}).

To understand the nature of the $\text{f}(R)$ black hole and to distinguish it from the known General Relativity
black holes, it is worthwhile to study their astrophysical features
such as the accretion of various kinds of fluids and their dynamics
near them. Using the isothermal and polytropic equations of state, we showed that the
standard method employed for tackling the accretion problem has masked
some important properties of the fluid flow.

Accretion of isothermal perfect fluids is is characterized by
\begin{itemize}
  \item Existence of subsonic flows for all values of the radial coordinate. These solutions represent neither
transonic nor supersonic flows as the fluid approaches the horizon;
  \item Existence of solutions with vanishing three velocity as the fluid approaches the horizon. As $v\to 0$, the fluid cumulates near the horizon resulting in a divergent pressure which pushes the fluid backward (flowout or a wind of the fluid under the effect of its own divergent pressure). These solutions, as the one depicted in Fig.~\ref{Fig3}, exist even in the case of a Schwarzschild black hole;
  \item If the CP is a saddle point, the critical solution curve divides the ($r,v$) plane into regions where the flow is physical in some of them (corresponding to higher values of the Hamiltonian) and unphysical in the others (corresponding to lower values of the Hamiltonian);
  \item The existence of separatrix heteroclinic orbits is subject to no constraint. We have checked this conclusion for the $\text{f}(R)$ model of Ref.~\cite{1e} and for Schwarzschild black hole and this should apply to all black holes;
  \item For the $\text{f}(R)$ model of Ref.~\cite{1e}, the existence of two CPs (one saddle and one center), with a possibly periodic flow inside a finite region of space, constraints the values of $\bt$ not to exceed some lower limit;
  \item Instability of the critical flow.
\end{itemize}

The polytropic test fluid has nearly no global solutions for the $\text{f}(R)$ model of Ref.~\cite{1e} unless one can deal with the fine tuning problem consisting in fixing the speed at spatial infinity in terms of the number density. Among the solutions we derived for the polytropic test fluid no saddle CP occurs. Moreover, the subsonic flow appears to be almost non-relativistic. This features appear quite different from the General Relativity black holes~\cite{CMS}.

de Sitter-like $\text{f}(R)$ black holes are characterized by the presence of closed, but non-homoclinic orbits, joining the event horizon to the cosmological horizon. Such cyclic curves are maintained by the high pressure present in the vicinity of the two horizons and do not require the presence of source-sink system for their realization. For $\ga>3/2$, the proper period of the cyclic flow converges to a finite value and has a logarithmically divergent limit for $\ga=3/2$. Comparison of the solutions (Figs.~\ref{Fig7} and~\ref{Fig8}) show that the accretion is insensitive to the $\text{f}(R)$ model.

\section*{Acknowledgment} We thank both anonymous reviewers for their very constructive comments and suggestions.

\section*{Appendix A: Roots of the Weierstrass polynomial\label{secaa}}
\renewcommand{\theequation}{A.\arabic{equation}}
\setcounter{equation}{0}

The Weierstrass polynomial is defined by
\begin{equation}\label{df1}
w(z)\equiv 4z^3-g_2z-g_3=4(z-e_1)(z-e_2)(z-e_3).
\end{equation}
Let $\De$ be the parameter
\begin{equation}\label{df2}
\De\equiv g_2^3-27g_3^2>0,
\end{equation}
the polynomial has the following properties~\cite{LP}.

\subsection{Three distinct real roots}
The Weierstrass polynomial $w(z)$ will have three real roots if
\begin{equation}\label{3.3}
    g_2>0 \quad\text{ and }\quad \De >0.
\end{equation}
We parameterize the (real) roots by the angle $0\leq \eta\leq \pi$ as follows~\cite{LP}:
\begin{align}\label{3.4}
&e_3=-\sqrt{\frac{g_2}{3}}\cos\Big(\frac{\pi-\eta}{3}\Big)<0,\quad
    e_2=-\sqrt{\frac{g_2}{3}}\cos\Big(\frac{\pi+\eta}{3}\Big),\nn\\
&e_1=\sqrt{\frac{g_2}{3}}\cos\Big(\frac{\eta}{3}\Big)>0,\\
&\cos\eta =\frac{9g_3}{\sqrt{3g_2^3}},\quad \sin\eta =\sqrt{\frac{\De}{g_2^3}}>0.\nn
\end{align}
With this parametrization it is obvious that $e_3<e_2<e_1$. The signs of $e_3<0$, $e_1>0$, and $\sin\eta>0$ are well defined,  and the sign of $e_2$ depends on that of $g_3$ ($g_3=4e_1e_2e_3$):
\begin{equation}\label{3.5}
    e_2 g_3<0\qquad (e_2=0\Leftrightarrow g_3=0).
\end{equation}

\subsection{Two distinct real roots}

The $w(z)$ will have two real roots if
\begin{equation}\label{3.12}
    g_2>0 \quad\text{ and }\quad \De =0.
\end{equation}
This happens when one of the local extreme values of $w(z)$ is zero.

\subsection{One real root}

The polynomial $w(z)$ will have one real root with multiplicity 1 if
\begin{equation}\label{3.23}
    \De <0.
\end{equation}
The sign of the real root $e_r$
\begin{equation}\label{4.23c}
    e_r=\frac{1}{2\cdot 9^{1/3}}[(9 g_3+ \sqrt{3} \sqrt{-\Delta })^{1/3}+(9 g_3-\sqrt{3} \sqrt{-\Delta })^{1/3}]
\end{equation}
is related to that of $g_3$ by
\begin{equation}\label{3.24}
    e_r g_3>0\qquad (e_r=0\Leftrightarrow g_3=0).
\end{equation}

\section*{Appendix B: Re-derivation of the critical points with  $\pmb{\mathcal{H}=\mathcal{H}(r,n)}$\label{secab}}
\renewcommand{\theequation}{B.\arabic{equation}}
\setcounter{equation}{0}

With $\mathcal{H}(r,n)$ given by~\eqref{h2}, the dynamical system reads
\begin{equation}\label{A1}
\dot{r}=\mathcal{H}_{,n}\,,  \quad\quad \dot{n}=-\mathcal{H}_{,r}.
\end{equation}
Evaluating the derivatives we obtain
\begin{align}
&\mathcal{H}_{,v}=2h^2\Big[\Big(f+\frac{C_1^2}{r^4n^2}\Big)(\ln h)_{,n}-\frac{C_1^2}{r^4n^3}\Big],\nn\\
\label{A2}&\mathcal{H}_{,r}=h^2\Big(f_{,r}-\frac{4C_1^2}{r^5n^2}\Big).
\end{align}
Using $(\ln h)_{,n}=a^2/n$~\eqref{19}, the system¨\eqref{A2} reads
\begin{align}
\label{A3a}&\dot{r}=\frac{2h^2}{r^4n^3}[a^2r^4n^2f+C_1^2(a^2-1)],\\
\label{A3b}&\dot{n}=-\frac{h^2}{r^5n^2}[r^5n^2f_{,r}-4C_1^2].
\end{align}
Setting the rhs's to zero we obtain
\begin{align}
\label{A4a}&a_c^2=\frac{C_1^2}{r_c^4n_c^2f+C_1^2},\\
\label{A4b}&f_{c,r_c}=\frac{4C_1^2}{r_cr_c^4n_c^2}.
\end{align}
Now, using~\eqref{v3} in~\eqref{A4a} and in~\eqref{A4b} we obtain $a_c^2=v_c^2$ and $r_c(1-v_c^2)f_{c,r_c}=4f_cv_c^2$, respectively. Since $a_c^2=v_c^2$, the equation $r_c(1-v_c^2)f_{c,r_c}=4f_cv_c^2$ is just the rightmost formula in~\eqref{cp1}.

For the other sonic point, $f_c=0$ and $a_c^2=1$, the rhs of~\eqref{A4a} is manifestly zero. The rhs of~\eqref{A4b} is also zero by~\eqref{v3} and~\eqref{cp1}. The latter provides the value of $f_{c,r_c}$ as the limit $r_c\to r_f$ and $a_c^2\to 1$.

\end{document}